\documentclass{jfm}

\usepackage{natbib}
\usepackage{graphicx}

\ifCUPmtlplainloaded \else
  \checkfont{eurm10}
  \iffontfound
    \IfFileExists{upmath.sty}
      {\typeout{^^JFound AMS Euler Roman fonts on the system,
                   using the 'upmath' package.^^J}%
       \usepackage{upmath}}
      {\typeout{^^JFound AMS Euler Roman fonts on the system, but you
                   dont seem to have the}%
       \typeout{'upmath' package installed. JFM.cls can take advantage
                 of these fonts,^^Jif you use 'upmath' package.^^J}%
      }
  \else
  \fi
\fi

\ifCUPmtlplainloaded \else
  \checkfont{msam10}
  \iffontfound
    \IfFileExists{amssymb.sty}
      {\typeout{^^JFound AMS Symbol fonts on the system, using the
                'amssymb' package.^^J}%
       \usepackage{amssymb}%
       \let\le=\leqslant  
       \let\ge=\geqslant  
      }{}
  \fi
\fi

\ifCUPmtlplainloaded \else
  \IfFileExists{amsbsy.sty}
    {\typeout{^^JFound the 'amsbsy' package on the system, using it.^^J}%
     \usepackage{amsbsy}}
    {\providecommand\boldsymbol[1]{\mbox{\boldmath $##1$}}}
\fi

\newsavebox{\astrutbox}
\sbox{\astrutbox}{\rule[-5pt]{0pt}{20pt}}

\def\gsim{\;\rlap{\lower 2.5pt
\hbox{$\sim$}}\raise 1.5pt\hbox{$>$}\;}
\def\lsim{\;\rlap{\lower 2.5pt
\hbox{$\sim$}}\raise 1.5pt\hbox{$<$}\;}
\newlength{\myfigwidth}
\usepackage{amsmath}

\title{The Pollution of Pristine Material in Compressible Turbulence}

\author[L. Pan, E. Scannapieco \& J. Scalo]{L\ls I\ls U\ls B\ls I\ls N\ns P\ls A\ls N$^1$\thanks{Email address for correspondence: liubin.pan@asu.edu}, 
\ns E\ls V\ls A\ls N\ns  S\ls C\ls A\ls N\ls N\ls A\ls P\ls I\ls E\ls C\ls O$^1$ \break \and J\ls O\ls H\ls N\ns S\ls C\ls A\ls L\ls O$^2$}

\affiliation{$^1$School of Earth and Space Exploration,  Arizona State University, P.O.  Box 871404, Tempe, AZ, 85287, USA\\[\affilskip]
${^2}$Department of Astronomy, University of Texas, Austin, TX 78712, USA}

\pubyear{??}
\volume{??}
\pagerange{??}
\date{?? and in revised form ??}
\begin{document}
\maketitle

\begin{abstract}



The first generation of stars had very different properties than 
later stellar generations, as they formed from a ``pristine" gas that 
was completely free of heavy elements. Normal star formation 
took place only after the first stars polluted the surrounding 
turbulent interstellar gas, increasing its local heavy element mass 
concentration, $Z$, beyond a  ``critical" threshold 
value, $Z_{\rm c}$ ($10^{-8} \lsim Z_{\rm c} \lsim 10^{-5}$).
Motivated by this astrophysical problem, we investigate the 
fundamental physics of the pollution of pristine fluid 
elements in statistically homogeneous and isotropic 
compressible turbulence. Turbulence stretches the 
pollutants, produces concentration structures at small scales, 
and brings the pollutants and the unpolluted flow in closer 
contact. The pristine material is polluted when 
exposed to the pollutant sources or the fluid elements polluted 
by previous mixing events. Our theoretical approach employs the 
probability distribution function (PDF) method for turbulent mixing, as the 
fraction of pristine mass corresponds to the low tail of the density-weighted concentration PDF.  
We adopt a number of PDF closure models and derive evolution equations 
for the pristine fraction from the models. To test and constrain the prediction
of theoretical models, we conduct numerical simulations for decaying 
passive scalars in isothermal turbulent  flows with 
Mach numbers of 0.9 and 6.2, and compute the 
mass fraction, $P(Z_{\rm c}, t)$, of the flow with $Z \le Z_{\rm c}$. 
In the Mach 0.9 flow,  the evolution of $P(Z_{\rm c}, t)$ is well 
described by a continuous convolution 
model and goes as $\dot P(Z_{\rm c}, t)= P(Z_{\rm c}, t) \ln[P(Z_{\rm c}, t)]/\tau_{\rm con}$, if the 
mass fraction of the polluted flow is larger than  $\approx 0.1.$ 
If the initial pollutant fraction is smaller than $\approx 0.1,$  
an early phase exists during which the pristine fraction follows 
an equation derived from a nonlinear integral model: 
$\dot P(Z_{\rm c}, t) =  P(Z_{\rm c}, t) [P(Z_{\rm c}, t)-1]/\tau_{\rm int}$.  
The timescales $\tau_{\rm con}$ and $\tau_{\rm int}$ are measured 
from our simulations. When normalized to the flow dynamical time, 
the decay of  $P(Z_{\rm c}, t)$ in the Mach 6.2 flow is slower than at 
Mach 0.9 because the timescale for scalar variance decay is slightly 
larger and the low tail of the concentration PDF broadens with 
increasing Mach number.  We show that  $P(Z_{\rm c}, t)$ in 
the Mach 6.2 flow can be well fit using a formula from a generalized version of the self-convolution model.  

\end{abstract}

\section{Introduction}

Big bang nucleosynthesis produced helium efficiently, but 
it was halted by the expansion of the universe before it 
was able to make stable elements heavier than lithium (Walker et al.\ 1991).
On the other hand, even the most pristine stars observed
today (Cayrel et al.\ 2004; Frebel et al.\ 2008; Caffau et al.\ 2011)
have substantial mass fractions of heavier elements, 
indicating that they have been polluted with the nucleosynthesis 
products of an as-yet undetected first generation of stars. 
This early stellar generation had an enormous impact on 
the evolution of later forming stars, and such stars are likely to have been much more massive 
(Abel, Bryan, \& Norman 2000; Bromm, Coppi \& Larson 2002) 
and much hotter (Schaerer 2002) than present-day stars, due to the important role that 
heavy elements play in star formation and evolution.

When and where this remarkable early stellar generation
formed is a question of fundamental astrophysical 
importance. On cosmological scales, the key issue 
is the time it takes for heavy elements to propagate 
from one galaxy to another. As shown in Scannapieco 
et al.\ (2003), the distances between these regions 
of early star-formation are so vast that the universe 
was divided into two regions: one in which galaxies 
formed out of material that was already polluted 
with heavy elements and one in which galaxies were 
formed from initially pristine material.

This second set of initially-pristine galaxies is 
especially interesting, as the first stars formed 
in these galaxies may be observable (Scannapieco 
et al.\ 2005; Jimenez \& Haiman 2006; 
Nagao et al.\ 2008). As star formation 
continued in these objects, the interstellar gas
became enriched with heavy elements released by the 
explosions of the first stars. This self-enrichment 
process increased the abundance or mass fraction, $Z$, 
of heavy elements, and finally led to a transition 
to normal star formation in regions where $Z$ 
exceeds a critical value, $Z_{\rm c}$. This critical 
value is expected to lie in the range $10^{-8} \lsim Z_{\rm c} \lsim10^{-5}$ 
(or $10^{-6}$ to $10^{-3}$ times the heavy-element abundance in the Sun), 
depending on whether the cooling of the 
interstellar gas is dominated by dust grains (Omukai et al.\ 2005) or by the fine 
structures of carbon and oxygen (Bromm \& Loeb 2003).

In a given galaxy, the key quantity to characterize 
the transition to normal star formation is the 
fraction, $P(Z_{\rm c}, t)$, of the interstellar gas 
with $Z$ below $Z_{\rm c}$ as a function of time. 
The temporal behavior of this fraction depends 
not only on the rate at which new sources of 
heavy elements are released to the interstellar gas, 
but, more importantly, on the transport and mixing 
process of these elements in the galaxy (Pan \& Scalo 2007). 
For example, a high mixing efficiency would result in a rapid 
decrease in $P(Z_{\rm c}, t)$, and hence in a sharp transition 
as the average concentration of heavy elements exceeds the 
threshold $Z_{\rm c}$. On the other hand, a low mixing 
efficiency would lead to a gradual transition. 
The interstellar gas in these galaxies is expected 
to be turbulent and highly compressible, and the 
turbulent motions are likely to be supersonic 
(Grief et al. 2008; Wise et al. 2008). Therefore, 
understanding  mixing in supersonic turbulence 
is crucial to answering the question of how the 
pristine gas in early galaxies was polluted.

In the present paper, we do not intend to directly model 
the complicated mixing process in a realistic galactic 
environment. Instead, we investigate the fundamental 
physics of turbulent mixing in compressible flows 
using idealized analytical and numerical tools. 
The primary goal is to understand the pollution of 
pristine material in statistically homogeneous and 
isotropic turbulence. This underlying physics is 
prerequisite for modeling the mixing of primordial gas 
in realistic interstellar turbulence. In a future work, we will 
apply the results of the current study to build a subgrid model for large-scale simulations 
for the formation and evolution of early galaxies. 
These simulations account for the complexities in the interstellar 
medium, but cannot resolve the scales at which true mixing occurs.
The subgrid model will provide a crucial step toward predicting 
the transition from primordial to normal star formation in the first generation 
of galaxies.    

A systematic numerical study of passive scalar physics 
in supersonic turbulence has been recently conducted 
by Pan and Scannapieco (2010, 2011), who simulated 
scalar evolution in six compressible turbulent flows 
with Mach number ranging from 1 to 6. In these papers, 
a detailed analysis of various statistical measures for 
the scalar field was performed, including the scalar 
dissipation, the scalar probability distribution, the 
power spectrum, the structure functions and intermittency. 
It was found that the classic cascade picture for passive 
scalars in incompressible turbulence is generally valid 
also for mixing in supersonic turbulent flows. The effect 
of compressible modes in supersonic turbulence and their modifications 
to the classic picture for passive scalar turbulence were 
examined by analyzing the Mach number dependence 
of the scalar statistics. The conclusions of these studies 
provide general theoretical guidelines for understanding 
the mixing process in interstellar turbulence. 

To explore how the pollutant-free mass is contaminated in 
turbulent flows, we make use of the probability distribution method 
for turbulent mixing.
The fraction of unpolluted or slightly polluted flow mass corresponds 
to the far left tail of the probability distribution function (PDF) of the 
concentration field, as $Z_{\rm c}$ is typically much smaller 
than the average value.  This fraction can be evaluated by integrating the 
concentration PDF from zero to the threshold, $Z_{\rm c}$. We will generally refer to the 
fraction $P(Z_{\rm c}, t)$ as the pristine fraction. Note that our approach here is general, and is not limited to mixing in early 
galaxies. 

The PDF equation for passive scalars cannot be solved exactly 
because of the closure problem, and various closure approximations have been 
developed to model the PDF evolution. In this work, we 
consider several existing closure models and derive equations for the fraction 
$P(Z_{\rm c}, t)$ for each of them. The 
far left PDF tail corresponds to high-order moments 
of the PDF, and thus it is quite uncertain whether the 
closure models can capture the high-order statistics with sufficient accuracy. 
In order to test the reliability of the adopted models and constrain 
their parameters, we perform numerical simulations for  turbulent mixing 
in a transonic flow and a highly supersonic flow. 

The structure of this paper is as follows. In \S 2, we present 
the general PDF formulation for mixing in compressible 
turbulence. \S 3 gives a brief description for several 
existing closure models for the diffusivity term in the PDF equation.  
The predictions of these models for the mass fraction 
of unpolluted or slightly polluted flow are derived in \S 4. 
We describe our numerical simulations in \S 5, which 
are used to test and constrain the theoretical models 
in \S 6. Our main conclusions are summarized in \S 7. 

\section{PDF Formulation for Mixing in Compressible Flows}

The PDF formulation was first developed for the probability 
distribution of the turbulent velocity field by Monin (1967) 
and Lundgren (1967), and for the PDF of the flow 
vorticity by Novikov (1967). The derivation of Monin (1967) 
was based on the equation for characteristic functions of 
the velocity field, while Lundgren (1967) started directly from 
the conservation laws of the flow. The two methods were 
later extended to derive PDF equations 
for scalar fields convected by a turbulent flow, such as the flow temperature or 
enthalpy, and the concentration fields of passive or reactive species in the flow 
(Ievlev 1973; Dopazo and O'Brien 1974; Pope 1976; O'Brien 1980; Pope 1985; 
Kollemann 1990; Dopazo et al.\ 1997). Recent discussions of PDF 
equations for passive or active 
scalar turbulence can be found in the monograph by Fox (2003) 
and the thorough reviews by Veynante \& Vervisch (2002) and Haworth 
(2010). 

In Appendix A, we derive the PDF equation for a passive 
scalar in compressible turbulence using the method 
of Lundgren (1967). The derivation is based on the 
equation of the concentration field, $C({\bf x}, t)$, for 
passive tracers advected in a turbulent flow with density $\rho({\bf x}, t)$ 
and velocity ${\bf v}({\bf x}, t)$: 
\begin{equation}
\frac{\partial  C} {\partial t}+   {\bf v}  \cdot {\bf \nabla} C =  \frac{1}{\rho} {\bf \nabla} \cdot (\rho \kappa {\bf \nabla} C) +  S(\bf{x},t),
\label{advection}
\end{equation}   
where the concentration field is defined as the 
ratio of the local tracer density to the flow density. 
In the diffusion term, $\kappa$ denotes the kinematic 
molecular diffusivity, and the dynamic diffusivity, $\rho \kappa$, 
is basically independent of $\rho$ (i.e., $\kappa \propto \rho^{-1}$). 
The term $S(\bf{x},t)$ represents the sources of new 
pollutants. 


Our derivation in Appendix A adopts a density-weighting scheme, which is 
appropriate for passive scalar mixing in compressible flows 
(Pan and Scannapieco 2010). We define a density-weighted 
concentration PDF, $p(Z; {\bf x}, t) \equiv \langle  \tilde{\rho} \delta[Z-C({\bf x}, t) ]\rangle$, 
where $\langle \cdot \cdot \cdot \rangle$ denotes the ensemble average, 
the density-weighting factor $\tilde{\rho}\equiv \rho({\bf x}, t)/\bar{\rho}$ 
is the ratio of the local flow density to the average density $\bar{\rho}$, 
and $Z$ is the sampling variable. Using the advection-diffusion 
equation (\ref{advection}), and the continuity equation for 
the evolution of the density-weighting factor, we obtain,  
\begin{equation}
\frac {\partial p (Z; {\bf x},t) }{\partial t} +  {\bf \nabla} \cdot \left(p \frac {\langle \rho {\bf v}|C=Z \rangle}
{\langle \rho |C=Z \rangle} \right)
= - \frac {\partial}{\partial Z} \left( p \frac {\langle \nabla \cdot (\rho \kappa \nabla C)|C=Z \rangle}{\langle \rho|C=Z \rangle} \right) -
\frac {\partial}{\partial Z} \left( p \frac {\langle \rho S|C=Z\rangle}  {\langle \rho|C=Z \rangle} \right), 
\label{pdfeq}
\end{equation}
where $\langle \cdot\cdot\cdot| C=Z \rangle $ denotes the ensemble 
average under the condition that the concentration field 
$C({\bf x}, t)$ is equal to $Z$ (see Appendix A). 
The equation is essentially a Liouville equation for the 
conservation of the concentration probability. To our 
knowledge, this equation for the scalar PDF with density-weighting 
has not been derived before.

The density weighting scheme is preferred in our study 
for two reasons. First, rather than the volume fraction, 
we are interested in the mass fraction of pristine gas 
in early galaxies, which corresponds to the left tail 
of the density-weighted PDF. Second, the advection 
term in the equation for the density-weighted PDF 
takes the form of a divergence, and thus conserves the 
global PDF (i.e., the integral of the local PDF, $p(Z; \bf{x},t)$, over 
the flow domain). This provides a formal and rigorous proof for the 
physical intuition that the turbulent velocity field itself does not 
homogenize the distribution of pollutants. The advecting 
velocity transports, redistributes and deforms the concentration field, 
but does not change the {\it mass} fraction of fluid elements with 
a given concentration level. Furthermore, the advection term 
vanishes if the flow and the concentration fluctuations 
are statistically homogeneous. 

In contrast, if one derives an equation for the volume-weighted 
PDF for a passive scalar in compressible turbulence, the 
advection term would not be a divergence term. The term 
reflects the effect of flow compressions and expansions, 
which can change the volume fraction of fluid elements 
at a given concentration (Pan \& Scannapieco 2010). 
This effect on the PDF is clearly different from scalar 
homogenization, and can be avoided by adopting a density-weighting 
factor. We thus argue that it is more appropriate to use the 
density-weighted PDF equation for the study of mixing in compressible turbulence.       

Molecular diffusion is the only process that homogenizes, 
and the molecular diffusivity term in the PDF equation 
continuously reduces the PDF width. This term can be rewritten as,  
\begin{equation}
- \frac {\partial}{\partial Z} \left( p \frac {\langle \nabla \cdot (\rho \kappa \nabla C)|C=Z \rangle}{\langle \rho|C=Z \rangle} \right)
=  - {\bf \nabla} \cdot \left[ \frac{\partial}{\partial Z} \left(p\frac{ \langle \rho \kappa \nabla C|C=Z \rangle} {\langle \rho|C=Z \rangle}\right) \right] 
-\frac{\partial^2}{\partial Z^2} \left(p \frac{\langle \rho \kappa ({\bf \nabla} C)^2 | C=Z\rangle} {\langle \rho | C=Z \rangle} \right),  
\label{diffterm}
\end{equation}
where both terms on the r.h.s depend on the ensemble 
average of the concentration gradients conditioned on $C({\bf x}, t) =Z$. 
As it is a divergence term, the first term in eq.\ (\ref{diffterm}) 
conserves the global concentration PDF. The scalar 
homogenization is achieved through the second term, which 
is essentially a diffusion term with a negative coefficient in 
concentration space. This term keeps narrowing the concentration PDF, 
and the physics of turbulent mixing can be viewed as 
an anti-diffusion process in concentration space.  

Taking the second order moment of the last term in eq.\ (\ref{diffterm}) 
gives the scalar dissipation rate, $- 2\langle \tilde{\rho} \kappa ({\bf \nabla} C)^2 \rangle$.  
Using this rate, we define a mixing timescale, 
\begin{equation}
\tau_{\rm m} \equiv \frac{\langle \tilde{\rho} \delta C^2 \rangle}{2 \langle \tilde{\rho}  \kappa ({\bf \nabla} C)^2 \rangle},
\label{mixingtime}
\end{equation} 
where $\delta C= C-\langle \tilde{\rho} C \rangle$ is the fluctuating 
part of the concentration field. The timescale, $\tau_{\rm m}$, 
corresponds to the scalar variance decay by mixing, and thus 
characterizes the rate at which the diffusivity term reduces the PDF width. 
Although the diffusivity terms in eqs.\ (\ref{diffterm}) and \ (\ref{mixingtime}) 
do not have an explicit dependence on the flow velocity, the 
mixing timescale is determined primarily by the turbulent velocity field. 
This is because the turbulent velocity produces progressively 
smaller structures and thus strongly amplifies the scalar 
gradients, $({\bf \nabla} C)^2$, in eqs.\ (\ref{diffterm}) and (\ref{mixingtime}). 
By feeding molecular diffusivity with large-gradient structures, 
turbulent motions greatly accelerate the scalar dissipation/homogenization. 

In the classic phenomenology for mixing in incompressible 
turbulence, the generation of small-scale concentration 
structures is through a cascade process similar to 
that of kinetic energy (Obukohov 1949; Corrsin 1951). The 
cascade is caused by continuous turbulent stretching, and 
it starts from the scale where the pollutant sources are injected into 
the flow, and proceeds to the diffusion scale where the molecular 
diffusion efficiently homogenizes the scalar fluctuations. 
The diffusion scale is essentially the scale 
where the action of molecular diffusivity becomes faster than turbulent 
stretching. From this picture, the mixing timescale, 
$\tau_{\rm m}$, is determined by the cascade time, which is essentially 
the eddy turnover time at the injection scale of the scalar 
sources because the cascade becomes faster and faster with decreasing length scale. 

Pan \& Scannapieco (2010) showed that the cascade picture also applies 
for mixing in supersonic turbulence. They found 
that the mixing timescale, $\tau_{\rm m}$, was close 
to the eddy turnover time at the pollutant injection scale 
in all their simulated flows with Mach numbers in the range 
from 1 to 6. 
The existence of compressible modes in supersonic 
flows causes only a slight Mach number dependence of the 
mixing timescale, and the primary ``mixer" is the stretching 
by solenoidal modes even at very high Mach numbers.   
   
Translating the physical discussion above to the mixing 
process of the unpolluted fluid elements in a turbulent flow 
gives the following picture. The 
turbulent velocity stretches the pollutants into smaller and 
smaller structures, and brings them to closer contact 
with the unpolluted flow. When the separation between 
the pollutant structures and the unpolluted fluid elements 
becomes close to or smaller than the diffusion scale, 
molecular diffusivity efficiently mixes them, 
reducing the unpolluted mass fraction. This 
suggests that the timescale for turbulent mixing to contaminate the 
unpolluted mass is also on the order of the scalar 
cascade timescale.   

The diffusivity term in the PDF equation has to be 
approximately modeled because of the closure problem 
(e.g., Dopazo and O'Brein 1974). Extensive efforts 
have been made to develop closure models for this term, 
and we will use several existing models 
in the current study, as described in \S 3. The advection term 
also has a closure problem, but modeling this term is not necessary if 
the flow and the scalar field are statistically homogeneous. 
The last term in the PDF equation (\ref{pdfeq}) 
corresponds to the effect of the pollutant sources. 
In reacting turbulent flows, the source term 
due to chemical reactions has a closed form in the 
PDF formulation (e.g., Pope 1976), and this has led to 
the wide use of the PDF method in studies of chemical 
reactions in turbulent flows. In the present study, 
the pollutant source is merely an initial scalar condition, 
and we do not discuss modeling the source term further 
(see Pan and Scalo 2007 for an example with a persistent source).  

\section{PDF Modeling}

\subsection{General Approach}

We employ both theoretical and numerical tools in the present work. 
The PDF formulation in \S 2 provides a general theoretical 
framework, and, to complete the theoretical approach, we will consider 
several existing closure models for the diffusivity term 
in eq.\ (\ref{pdfeq}).  We will compare the predictions of these models for 
the scalar PDF evolution with that measured from numerical 
simulations. From our simulation data, we compute the 
concentration PDF as $p(Z; t) = \frac{1}{V} \int_V \tilde{\rho} \delta [Z-C({\bf x}, t)]   d{\bf x}$, 
where $V$ is the total volume of the simulation box. This PDF measures 
the concentration fluctuations over the entire flow domain, and thus should 
be viewed as a global PDF. As pointed out in \S 2, the advection term 
conserves the global density-weighted PDF, and thus need not be 
considered in our tests of the theoretical models against simulations. The global 
PDF is expected to be equal to the local PDF, $p(Z; {\bf x}, t)$, defined in 
the ensemble context under the assumption of statistical homogeneity. 

In our simulations, we only evolve decaying scalars with the source 
term $S({\bf x}, t)$ set to be zero. Neglecting the advection and 
source terms, the PDF equation becomes   
\begin{equation}
\frac {\partial p (Z;t) }{\partial t} = - \frac {\partial}{\partial Z} \left( p \frac {\langle \nabla \cdot (\rho \kappa \nabla C)|C=Z \rangle}{\langle \rho|C=Z \rangle} \right). 
\label{simplepdfeq}
\end{equation}
The only term that contributes to the PDF evolution in 
our simulations is the diffusivity term, and modeling this term 
is the main task of the PDF approach to turbulent mixing. 
In incompressible turbulence, the flow density is 
constant, and eq.\ (\ref{simplepdfeq}) reduces to,  
\begin{equation}
\frac {\partial p (Z; t) }{\partial t} = - \frac {\partial}{\partial Z} \left( p \langle \kappa \nabla^2 C)|C=Z \rangle \right),    
\label{incomppdfeq}
\end{equation}
which has been extensively  studied and modeled.  

The second order moment of eq.\ (\ref{simplepdfeq}) corresponds to the scalar variance equation,   
\begin{equation}
\frac{d\langle \delta Z^2 \rangle}{dt} = - \frac{\langle \delta Z^2 \rangle}{\tau_{\rm m}},
\label{variance}
\end{equation}
where $\langle \delta Z^2 \rangle \equiv \langle \tilde{\rho} \delta C^2 \rangle$ 
denotes the density-weighted variance, and we have used the 
definition, eq.\ (\ref{mixingtime}), of the mixing timescale. 
In terms of the PDF, the variance is given by $\langle \delta Z^2 \rangle = \int (Z-\bar{Z})^2 p(Z, t) dZ$ 
with $\bar{Z} = \int Z p(Z; t)dZ $ being the mean concentration. In general, 
$\tau_{\rm m}$ may be a function of time. But if it is constant, the scalar variance 
decreases exponentially, which is the case at the late evolution stage of a 
decaying scalar (see \S 6.3). 

In analogy to the enrichment of pristine gas by the first 
generation of stars, the initial condition of the decaying 
scalars in this study will be set to be bimodal: consisting of pure pollutants 
($Z=1$) and completely unpolluted flow ($Z=0$). This corresponds to a 
double delta function form for the initial concentration PDF,
\begin{equation}  
p(Z; 0) = P_0 \delta(Z) +  P_1 \delta(Z-1),  
\label{initialpdf}
\end{equation}
where $P_1$ and $P_0$ are the initial mass fractions of 
the pollutants  and the unpolluted flow, respectively, 
and we have $P_0 +P_1 =1$ from the normalization of 
the PDF.  


The rest of this section is devoted to modeling the 
diffusivity term in the PDF equation. A variety of 
closure models have been proposed for this term, and 
the interested reader is referred to Dopazo et al.\ (1997) 
and Haworth (2010) for reviews. Here we will consider 
three of the models proposed for the diffusivity closure:  
the mapping closure model by Chen at al.\ (1989), the nonlinear 
integral models by Curl (1963), Dopazo (1979) and 
Janicka et al.\ (1979), and the self-convolution models by Villermaux \& 
Duplat (2003), Venaille and Sommeria (2007) and Duplat \& Villermaux (2008).   

We point out that, in compressible turbulence, the 
diffusivity term has an explicit dependence on the density field, 
or more precisely, on the joint statistics of the density 
and the concentration fields. Therefore, an ideal PDF 
model for mixing in supersonic turbulence needs to 
account for the effect of density fluctuations 
on the diffusivity term, and to predict the dependence 
of the concentration PDF on the flow compressibility. However, 
to our knowledge, this has not been considered in existing models, 
which were usually tested against simulation results for mixing 
in incompressible turbulence. We will compare the predictions 
of the closure models mentioned above with our simulation 
data, and examine whether, by adjusting their parameters, 
these models can be successfully applied to the contamination 
process of pollutant-free mass in compressible turbulent flows 
at different Mach numbers. Future studies are motivated to develop 
closure models that explicitly address the effects of shocks 
and the Mach number dependence of the 
passive scalar PDF in supersonic turbulence. 

\subsection{The Mapping Closure Model}

We first discuss the mapping closure model developed by Chen et al.\ (1989) 
for mixing in incompressible turbulence. We give a brief introduction 
to the model, and a detailed derivation can be found in, 
e.g., Pope (1991). The model is based on a surrogate field, 
$\phi ({\bf x}, t)$, obtained from the mapping of a Gaussian reference field $\theta({\bf x}, t)$,   
\begin{equation}
\phi({\bf x}, t) = X[\theta({\bf x}, t), t], 
\label{mapping}
\end{equation} 
where $X$ is an ordinary, non-stochastic function. The Gaussian 
field, $\theta({\bf x}, t)$, is assumed to be statistically 
homogeneous and have zero mean and unit variance, i.e., 
the probability of finding $\theta({\bf x}, t)$ equal to a given value $\eta$ is 
given by $g(\eta) \equiv \frac{1}{\sqrt{2 \pi}} \exp(-\eta^2/2)$ 
(see Girimaji (1992) for a generalized version of the mapping 
closure where the reference field PDF is time-evolving and not limited to Gaussian).  
The main idea of the model is to pursue a mapping function, 
$X(\eta, t)$, with which the PDF of the surrogate field obeys 
exactly the same equation [i.e., eq.\ (\ref{incomppdfeq})] as the 
actual field, $C({\bf x}, t)$. This is indeed achieved if the mapping 
function evolves as
\begin{equation}
\frac {\partial X(\eta, t)}{\partial t} = \frac{\kappa}{\lambda^2_\theta(t)} \left(\frac{\partial^2 X(\eta, t)}{\partial \eta^2} - \eta \frac{\partial X(\eta, t)}{\partial \eta} \right),
\label{mappingeq}
\end{equation}
where $\lambda_\theta(t) = \langle ({\bf \nabla} \theta)^2 \rangle^{-1/2}$, 
and ${\lambda^2_\theta(t)}/\kappa$ is a timescale that controls the 
rate at which the mapping function and hence the PDF evolve. 
The timescale is unspecified in the original model, and can be 
calibrated by a comparison of the variance decay in the model 
with simulation results (see He and Zhang (2004) 
for a theoretical evaluation of this timescale using a two-point 
closure strategy).  The evolution equation for the mapping function 
was derived in the incompressible limit, and thus 
the model is intended for mixing in incompressible 
turbulence only. By a comparison with our simulation 
data, we will examine whether the mapping closure model 
may also give acceptable predictions for mixing in compressible turbulence.   
   
With the desired mapping function, one can 
approximate the PDF of the actual field with that of 
the surrogate field. Using eq.\ (\ref{mappingeq}), the PDF of 
the surrogate field can be converted from the 
Gaussian PDF of the reference field. The conversion gives, 
\begin{equation}
p(Z; t) = g(\eta) \left( \frac{\partial X (\eta, t)}{\partial \eta} \right)^{-1}, 
\label{predictedpdf}
\end{equation} 
where $\eta$ is the solution of $X(\eta, t) =Z$. 

The linear equation for the mapping function, eq.\ (\ref{mappingeq}), 
can be solved analytically, provided the initial condition $X(\eta, 0)$ (Pope 1991). 
For a double-delta initial PDF, the initial mapping is a Heaviside 
step function $X(\eta, 0) = H(\eta-\eta_0)$, where $\eta_0$ 
satisfies $\int_{-\infty}^{\eta_0} g(\eta) d\eta = P_0$. With this 
initial condition, $X(\eta, t)$ is solved by 
\begin{equation}
X(\eta, t) = G\left(\frac{\eta}{\Sigma(t)} - \frac{\eta_0 (\Sigma(t)^2 +1)^{1/2}}{\Sigma(t)} \right),
\label{pdfsolution}
\end{equation} 
where $\Sigma(t)^2 = \exp [\int_0^t \kappa/\lambda^2_\theta(t') dt'] -1$,  
and $G(\eta) \equiv \int_{-\infty}^{\eta} g(\eta') d\eta'$ is the cumulative 
function of the Gaussian function. Combining eqs.\ (\ref{predictedpdf}) 
and (\ref{pdfsolution}) gives the predicted PDF evolution by the 
mapping closure model.


\subsection{The Nonlinear Integral Models}

In this subsection, we consider a class of closure models that 
use an integral form to approximate the diffusivity term in the 
PDF equation. This type of models originates from the 
equation introduced by Curl (1963),     
\begin{equation}
\frac{\partial p(Z; t)}{\partial t} = \gamma(t) \left\{ \left[ \int\limits_{0}^{1} dZ_{1}  p(Z_{1};t) 
\int\limits_{0}^{1} dZ_{2} p(Z_{2}; t) \delta \left( Z-\frac{Z_{1} +Z_{2}}{2} \right) \right]  -p(Z; t) \right\},  
\label{curl}
\end{equation}
where $\gamma(t)$ is the turbulent stretching rate.  A physical interpretation 
of this equation is as follows (e.g., Dopazo 1979). The turbulent velocity field 
stretches the concentration field and produces structures at small scales. As 
shown by various studies, these structures are primarily in the form of 2D sheets 
(e.g., Pan and Scannapieco 2011 and references therein). The scalar 
sheets are brought closer to each other over time by the turbulent velocity. 
When the typical width and separation of the sheets decrease to 
the diffusion scale, molecular diffusivity can operate efficiently and 
homogenize. The timescale for this process is $\gamma(t)^{-1}$, which is 
expected to be on the order of the scalar cascade timescale or the mixing 
timescale $\tau_{\rm m}$. Two scalar sheets of different concentrations brought to close contact are assumed to mix perfectly, 
resulting in a concentration value equal to their average prior to the mixing event 
[see the delta function in eq.\ (\ref {curl})]. The last term in eq.\ (\ref {curl}) 
corresponds to the ``destruction" of the previous PDF by the mixing event. 

One problem of Curl's model for turbulent mixing is that, if the initial 
concentration PDF consists of two delta functions [see eq.\ (\ref{initialpdf})], 
the predicted PDF shows unphysical spikes in between the initial delta 
functions. To avoid this problem, Dopazo (1979) and Janicka et al.\ (1979) 
independently generalized Curl's model replacing the delta function in eq.\ (\ref{curl}) 
by a smooth function $J(Z; Z_{1}, Z_{2})$, 
\begin{equation}
\frac{\partial p(Z; t)}{\partial t} = \gamma(t) \left\{ \left[ \int\limits_{0}^{1}p(Z_{1};t) \int\limits_{0}^{1} p(Z_{2}; t)J(Z; Z_{1}, Z_{2}) dZ_{1} dZ_{2} \right]  - p(Z; t) \right\},  
\label{dopjangeneral}
\end{equation}
where $J(Z; Z_{1}, Z_{2})$ represents the effect of mixing between two nearby scalar 
sheets with concentration values of $Z_1$ and $Z_2$. The 
function $J(Z; Z_{1}, Z_{2})$ is zero for $Z$ outside the range $(Z_1, Z_2)$ (or $(Z_2, Z_1)$ if $Z_1 > Z_2$), 
and its normalization is $\int_{Z_{1}}^{Z_{2}} J(Z; Z_1, Z_2) dZ =1$. 
A simple assumption for $J(Z; Z_{1}, Z_{2})$ is that it is uniform between $Z_1$ and $Z_2$, leading to  
\begin{equation}
\frac{\partial p(Z; t)}{\partial t} = \gamma(t)\left\{ \left[  \int\limits_{0}^{Z} p(Z_{1};t) \int\limits_{Z}^{1}  \frac{2}{Z_2 - Z_1}  p(Z_{2}; t)  dZ_1 dZ_2 \right] -p(Z; t) \right\}, 
\label{dopjan}
\end{equation}
where  $J(Z; Z_{1}, Z_{2})$ was set to $1/|Z_2-Z_1|$ (Dopazo 1979 and Janicka et al.\ 1979). 

The parameter $\gamma(t)$ as a function of time can be fixed by comparing 
the variance equation of these models with the simulation data.  The derivation 
for the variance equation can be found in Janicka et al.\ (1979) or Valino 
and Dopazo (1990). For Curl's model and the model with uniform $J(Z; Z_{1}, Z_{2})$, 
the variance decays as $\propto \exp[- \frac{1}{2}\int_0^t \gamma(t') dt']$ and 
$\exp[- \frac{1}{3}\int_0^t \gamma(t') dt']$, respectively. If the variance decreases 
exponentially with a constant timescale $\tau_{\rm m}$, $\gamma(t)$ 
is constant and equal to $2/\tau_{\rm m}$ and $3/\tau_{\rm m}$, 
respectively, for the two models. 
  
Pope (1982) pointed out a weakness of this class of models: the normalized 
high-order moments, $\langle \delta Z^m \rangle/ \langle \delta Z^2\rangle^{m/2}$, 
do not converge with time for $m \ge 4$ (see also Valino and Dopazo 1990). 
This suggests that the predicted PDF by these models has excessively 
fat tails at late times (Kollemann 1990). 

\subsection{The Self-convolution Models} 

The last type of models we consider are those based on the 
self-convolution of the scalar PDF, which can be viewed as extensions 
of the model by Curl (1963) in Laplace space. A review for the development 
of these models can found in Duplat and Villermaux (2008). 

The Laplace transform $\hat{p}(\zeta; t)$ of the scalar PDF is defined as 
$\hat{p}(\zeta; t) = \int_0^{\infty} p(Z; t) \exp(-Z \zeta) dZ$. Using the 
convolution theorem, the Laplace transform of eq.\ (\ref{curl}) gives,  
\begin{equation}
\frac{\partial \hat{p} (\zeta; t)}{\partial t} =\gamma \left[\hat{p} (\zeta/2; t)^2 - \hat{p} (\zeta; t)\right].  
\label{convolution}
\end{equation}
A similar equation in Fourier space was used by Pumir et al.\ (1991). 
This equation shows that turbulent mixing is essentially 
treated as a self-convolution process in Curl's model. We rewrite eq.\ (\ref{convolution}) 
in a difference form $\hat{p} (\zeta; t+ \delta t) = \epsilon \hat{p} (\zeta/2; t)^2 +(1-\epsilon) \hat{p} (\zeta; t)$
where $\epsilon = \gamma \delta t$ with $\delta t$ an infinitesimal time 
step. The difference equation can be interpreted as: during a time step $\delta t$, 
mixing occurs only in an infinitesimal fraction, $\epsilon$, of the flow, and 
in this part of the flow the scalar PDF undergoes a complete convolution.  The 
convolution process in eq.\ (\ref{convolution}) appears to be ``discrete". 

Following Venaille and Sommeria (2007), we derive a continuous 
version of Curl's model. We assume that, in each time step $\delta t$, 
the PDF convolution occurs everywhere in the flow, but the number 
of convolutions is taken to be infinitesimal and equal to $\epsilon$ 
(Duplat and Villermaux 2008). This assumption can be written as $\hat{p} (\zeta; t+ \delta t) = \hat{p} (\zeta/(1+\epsilon); t)^{(1+\epsilon)}$. 
Using the Taylor expansion $\hat{p} (\zeta/(1+\epsilon); t)^{(1+\epsilon)} 
\simeq \hat{p} (\zeta; t) +  \epsilon [\hat{p}(\zeta; t) \ln (\hat{p}(\zeta; t) ) -\zeta \partial \hat{p}(\zeta; t)/\partial \zeta]$ 
and taking the limit $\delta t \to 0$, we obtain, 
\begin{equation}
\frac{\partial \hat{p} (\zeta; t)}{\partial t} = \gamma \left[ \hat{p} \ln (\hat{p}) -  \zeta \frac {\partial \hat{p}}{\partial \zeta} \right], 
\label{cconvolution}
\end{equation} 
which represents the model of Venaille and Sommeria (2007). We will refer to this model 
as the continuous convolution model.  In this model, the variance decays as $\propto \exp(-\int_0^t \gamma(t') dt')$.
Venaille and Sommeria (2007) showed that  the predicted PDF by 
eq.\ (\ref{cconvolution}) evolves toward Gaussian in the long time limit (in contrast to the integral 
models in \S 3.3). A comparison of this model with experimental data is given in 
Venaille and Sommeria (2008).  We note that, if the initial PDF is two delta 
functions, the continuous self-convolution model is not applicable for the PDF evolution 
right from the beginning (Venaille and Sommeria 2007). We thus cannot compare 
the model prediction for $p(Z; t)$ with our simulation results at the early evolution stage. 
The model will only be used to study the evolution of the unpolluted mass fraction. 

As pointed out by Duplat and Villermaux (2008), a more general extension of Curl's model is,  
\begin{equation}
\frac{\partial \hat{p} (\zeta; t)}{\partial t} = \gamma n \left[ \hat{p} (\frac{\zeta}{1+1/n}; t)^{(1+1/n)} - \hat{p} (\zeta; t) \right]. 
\label{nconvolution}
\end{equation} 
Curl's original model (eq.\ (\ref{convolution})) and the model of Venaille and Sommeria (2007) 
(eq.\ (\ref{cconvolution})) are special cases of eq.\ (\ref{nconvolution}) with $n=1$ and $n \to \infty$, 
respectively. The parameter $n$ can be a function of time in general.  
The assumption behind eq.\ (\ref{nconvolution}) is that a fraction, $n \epsilon$, of the flow 
experiences mixing/convolution events during a time step $\delta t$, and the number of 
convolutions in this fraction of the flow is $1/n$.  For eq.\ (\ref{nconvolution}), the variance decay goes 
like $\propto \exp(-\int_0^t \gamma(t') n/(n+1) dt')$. We will refer to  eq.\ (\ref{nconvolution}) as the 
generalized convolution model.  

We finally consider the model by Villermaux and Duplat (2003), which was motivated 
by a turbulent mixing picture with three related processes: the generation of pollutant 
sheets by turbulent stretching, the diffusion of the pollutant sheets by molecular diffusivity 
and the merging of the diffused sheets. The merging of the sheets corresponds to 
a self-convolution process. The model is represented by (Duplat and Villermaux 2008), 
\begin{gather}
\frac{\partial \hat{p} (\zeta; t)}{\partial t} = - \gamma \zeta \frac {\partial \hat{p}}{\partial \zeta}  + n \gamma \left[\hat{p} (\zeta; t)^{(1+1/n)} - \hat{p} (\zeta; t) \right], \notag \\
\frac{\partial n(t)}{\partial t} = \gamma n, 
\label{aggregation}
\end{gather} 
where $n$ increases with time and the first equation can be viewed as 
the expansion of eq.\ (\ref{nconvolution}) at large $n$. Note that eqs.\ (\ref{aggregation}) 
and (\ref{cconvolution}) approach the same limit as $t \to \infty$. Villermaux and Duplat (2003) 
showed that eq.\ (\ref{aggregation}) has an asymptotic solution $\hat{p} (\zeta; t) = (1+ \langle Z \rangle \frac{\zeta}{n})^{-n}$ at large 
$t$, which corresponds to a Gamma distribution for the scalar 
PDF (Duplat and Villermaux 2008), 
\begin{equation}
p(Z; t) =  \frac{n^n}{\Gamma(n)\langle Z\rangle^n} Z^{n-1} \exp \left(-\frac{nZ}{\langle Z\rangle}\right),  
\label{gamma}
\end{equation}
where $\Gamma(n)$ is the Gamma function. The Gamma 
distribution is valid only at late times with $n \gsim1$, 
and cannot be applied to study the pristine mass fraction 
at the early evolution stage when the fraction is significant. 
Therefore, we do not use the model for the pristine mass 
fraction, but will check whether the scalar PDF in 
our simulations approaches a Gamma distribution at 
late times.     

We point out a fundamental difference between the mapping 
closure model discussed in \S 3.2 and the models presented here 
and in \S 3.3. The mapping closure is established by a 
direct approximations of the exact, but unclosed form of the diffusivity term. 
On the other hand, the nonlinear integral models and the convolution models 
do not start from the diffusivity term in the PDF equation, instead they are largely 
based on a physical picture for the mixing process. 

\section{Mass Fraction of Unpolluted or Slightly Polluted Flow}

As mentioned in the Introduction, we are interested 
in the mass fraction, $P(Z_{\rm c}, t)$, of the flow 
with concentration smaller than a tiny threshold, 
$Z_{\rm c}$, which can be calculated from the 
concentration PDF as  
\begin{equation}
P( Z_{\rm c}, t) = \int\limits_0^{Z_{\rm c}} p(Z'; t) dZ'. 
\label{fraction}
\end{equation} 
The fraction corresponds to the far left tail of the 
PDF since the threshold $Z_{\rm c}$ of interest 
is typically much smaller than the average 
concentration. 
Taking the limit $Z_{\rm c} \to 0$ in eq.\ (\ref{fraction}), 
we obtain the fraction $P(t)$ of exactly pollutant-free 
mass, i.e., $P( t) = \lim_{Z_{\rm c} \to 0 } \int_0^{Z_{\rm c}} p(Z'; t) dZ'$. 
This fraction is zero unless $p(Z; t)$ has a delta function 
component, $\delta (Z)$, at $Z=0$. In this section, we derive 
equations for $P(Z_{\rm c}, t)$ and $P(t)$ from the closure models 
discussed in \S 3. 

An interesting observation of the action of molecular 
diffusivity is that it tends to decrease the exactly 
pollutant-free fraction, $P(t)$, to zero instantaneously. 
For illustration, we consider a simple situation with a point source 
diffusing in a static uniform medium. The concentration field obeys 
the diffusion equation, whose solution is given by a 
Gaussian function. From this solution, it is clear that, 
no matter how small the molecular diffusivity, $\kappa$, is, 
the concentration field at a finite time ($t>0$) becomes nonzero at 
any finite distance ($r < \infty$) from the initial source, 
suggesting that all the pollutant-free mass is 
removed from the system instantaneously.  

This acausal behavior of molecular diffusivity 
originates from the Laplacian operator in the 
diffusion equation, which implicitly assumes that 
the random walk of some tracer molecules 
can bring them to an infinite distance 
during any small (but macroscopic) time interval. 
This is clearly unrealistic. The thermal motions 
of tracer molecules must have a finite 
maximum speed, $\max(v_{\rm th})$, and thus none of 
them can reach an infinite distance instantaneously. 
If the size of the system in question is $L$, there could be 
exactly pollutant-free mass surviving for a finite time 
$\sim L/ \max(v_{\rm th})$. However, this time is 
expected to be very small since $\max(v_{\rm th})$ is 
likely to much larger than the sound speed. Therefore, 
the reduction of exactly pollutant-free fraction, $P(t)$, 
by molecular diffusion may be considered as being 
essentially instantaneous.   

For our astrophysical applications, we need  the fraction, 
$P(Z_{\rm c}, t)$, of the flow with $Z$ below a finite 
critical value, $Z_{\rm c}$, rather than the exactly 
pristine fraction. Obviously, it takes finite time for molecular diffusivity to enrich all 
the fluid elements in the system up to a finite threshold, 
$Z_{\rm c}$. In fact, during a short time interval, the 
degree of pollution by molecular diffusivity alone is 
negligible even at small distances from the pollutant 
source, and the entire system is practically unpolluted. 
Therefore, the observation of the rapid/immediate 
erasure of exactly pristine gas by molecular diffusivity 
is not directly relevant to the astrophysical problem 
of primordial star formation.   

Because $\kappa$ is usually tiny in practical 
environments, such as in the interstellar 
media of galaxies, enriching 
all the fluid elements to a concentration level of, 
say, $\gsim 10^{-8}$, by the molecular diffusivity alone is very 
slow (see discussions in Pan and Scalo 2007). 
The presence of a turbulent velocity field greatly accelerates 
the mixing process, making the reduction of $P(Z_{\rm c}, t)$ 
much faster.  We find that the timescale for the 
reduction of $P(Z_{\rm c}, t)$ with a small $Z_{\rm c}$
is basically determined by the rate at which the 
turbulent stretching produces small-scale 
structures and is essentially independent of $\kappa$. 
            
\subsection{The Mapping Closure Model}

We calculate the fraction $P(Z_{\rm c}, t)$ predicted 
by the mapping closure model. From eq.\ (\ref{predictedpdf}), it is 
straightforward to find that    
\begin{equation}
P(Z_{\rm c}, t) = \int\limits_{-\infty}^{\eta_{\rm c} (t)} g(\eta) d\eta = G (\eta_{\rm c} (t)),
\end{equation} 
where the upper limit $\eta_{\rm c}(t)$ satisfies $X(\eta_{\rm c}(t); t) = Z_{\rm c}$. 
For a given value of $Z_{\rm c}$, the limit  $\eta_{\rm c}(t)$ changes with 
time as the mapping function evolves, and for our initial bimodal 
PDF with two delta functions, $\eta_{\rm c}(t)$ can be computed using  eq.\ (\ref{pdfsolution}).   

From that equation, we see that $Z_{\rm c}=0$ corresponds 
to $\eta_{\rm c} \to -\infty$ at all times after $t=0$.  Therefore, 
the mapping closure model predicts that $P(t)$ is zero at 
any time $t>0$, or that  the fraction of exactly pollutant-free 
mass deceases to zero instantaneously. This is consistent 
with our discussion above that the molecular diffusivity 
alone tends to immediately remove fluid elements with 
exactly zero concentration. The mapping closure model 
inherits this particular property of molecular diffusion, because 
the effect of diffusivity as a Laplacian term is treated directly. 
The model destroys the initial delta function at $Z=0$ 
instantaneously. However, this does not suggest that $p(0; t)$ 
becomes finite immediately. At the early evolution stage, $p(Z; t)$ 
does have an infinite peak at $Z=0$, but the peak is less 
singular than a delta function ($\delta(Z)$) in the sense 
that $\int_0^Z p(Z'; t)dZ' \to 0$ in the limit $Z \to 0$. 

\subsection{The Nonlinear Integral Models}

Unlike the mapping closure model, the nonlinear 
integral models preserve the singularities at $Z=0$ 
and $Z=1$. More specifically, the amplitudes of the delta functions 
at $Z=0$ and $Z=1$ decrease with time, but they are never 
completely destroyed, such that exactly pollutant-free mass 
can survive in these models, and $P(t)$ remains finite at any finite time. This 
is inconsistent with our earlier observation that the 
molecular diffusivity tends to reduces $P(t)$ to zero 
immediately. The reason is that the effect of molecular 
diffusivity is not incorporated directly in these models, instead it 
is included implicitly through the function $J(Z; Z_1, Z_2)$. 
Despite the inconsistency, we find that the integral models 
are useful for understanding the pollution of fluid 
elements with very low (but nonzero) concentration 
by turbulent mixing. Below we derive an equation 
for the fraction, $P(t)$, of exactly pollutant-free mass from these 
models.  

We consider the general model represented by eq.\ (\ref{dopjangeneral}). 
Integrating this equation in the range $[0, Z]$ and taking the limit $Z \to 0$, we have, 
\begin{equation}
\frac{dP(t)}{dt}= \gamma(t) \left( \int\limits_{0}^{1} dZ_1 p(Z_{1};t) \int\limits_{0}^{1} dZ_2  p(Z_2; t) \lim_{Z \to 0} \int\limits_{0}^{Z} dZ' J(Z'; Z_1, Z_2)  -P(t) \right).  
\label{dopjanpf}
\end{equation}
The last integral in the triple-integral term in the limit $Z \to 0$ can be written 
as $\int_{0}^{0^+} J(Z'; Z_1, Z_2) dZ'$ where $0^+$ represents the upper integral 
limit approaching zero from the positive vicinity. We first note that this 
integral is zero if both $Z_1$ and $Z_2$ are positive because  $J(0; Z_1, Z_2) = 0$ 
for $Z_1>0$ and $Z_2>0$ (see \S 3.3). We next assume that $J(Z; Z_1, Z_2)$ at $Z=Z_1$ 
and $Z=Z_2$ is nonsingular or less singular than a delta function 
for $Z_1 \ne Z_2$ (meaning that $\int_{Z_1}^{Z_1^+} J(Z; Z_1,Z_2) dZ =0$ 
and $\int_{Z_2^-}^{Z_2} dJ(Z; Z_1,Z_2) dZ =0$, where $Z_1 <Z_2$ is 
assumed without loss of generality). This assumption is clearly satisfied for Curl's 
model and the model with uniform $J(Z; Z_{1}, Z_{2})$ (eq.\ (\ref{dopjan})).  
With this assumption, it is straightforward to see that $\int_{0}^{0^+} J(Z'; Z_1, Z_2) dZ'$ 
is finite only if both $Z_1=0$ and $Z_2=0$. In that case, we have $\int_{0}^{0^+} dZ' J(Z'; 0, 0)=1$ 
from the normalization of $J(Z; Z_1, Z_2)$. This observation suggests that 
the contribution to the triple integral in eq.\ (\ref{dopjanpf}) comes only from 
$Z_1$ and $Z_2$ values in an infinitesimal range around zero. With 
such infinitesimal ranges of $Z_1$ and $Z_2$, the first two of the three 
integrals contribute factors of $\int_{0}^{0^+} p(Z_{1};t) dZ_1$ and 
$\int_{0}^{0^+} p(Z_{2};t) dZ_2$, respectively. As both these factors are 
equal to $P(t)$,  we have the following equation for $P(t)$,  
\begin{equation}
\frac{dP(t)}{dt}= -\frac{P(1-P)}{\tau_{\rm int}}, 
\label{pfintegral}
\end{equation}  
where $\tau_{\rm int} \equiv  \gamma(t)^{-1}$ is used for the convenience 
of notations. If the mixing timescale $\tau_{\rm m}$ is constant, the variance 
decay requirement gives $\tau_{\rm int} =  \tau_{\rm m}/2$ or 
$\tau_{\rm m}/3$ for Curl's model and the model with 
uniform $J(Z; Z_{1}, Z_{2})$, respectively (see \S 3.3). 

The equation gives an interesting physical picture for mixing 
of the unpolluted mass in turbulent flows: the pristine 
fraction is reduced when turbulent stretching brings 
the pollutant-free fluid elements (with a fraction of $P(t)$),
and the rest of the flow (with a fraction of $1-P(t)$), which 
has been polluted by sources or previous mixing events, close 
enough for molecular diffusivity to homogenize (Pan and Scalo 2007). 

Eq.\ (\ref{pfintegral}) has a simple analytic solution,   
\begin{equation}
P(t) = \frac{P_0}{P_0 + (1-P_0) \exp \left( \frac{t}{\tau_{\rm int}} \right)},  
\label{pfintegralsolution}
\end{equation}   
where $P_0$ is the initial fraction of unpolluted mass, and we 
have assumed $\tau_{\rm int}$  is constant with time. 
Although it is derived for the fraction of exactly pollutant-free 
mass, we will show in \S 5 that, in certain physical 
regimes, this equation can be used to fit our numerical 
results for $P(Z_{\rm c}, t)$ with a finite threshold $Z_{\rm c}$. 
        
\subsection{The Self-convolution Models}  

The self-convolution models introduced in \S 3.4 also preserve 
the initial singularities at $Z=0$ and $Z=1$, since they are 
essentially extensions of Curl's model. Again we derive the equations for the 
fraction, $P(t)$, of exactly pollutant-free mass from 
the convolution models,  which will be used later to 
understand the mass fraction of nearly-pristine, but $Z \neq 0$, flow.
We first decompose the concentration PDF as  
\begin{equation}
p(Z;t) =P(t) \delta (Z) + p_{\rm e}(Z; t), 
\label{decomp}
\end{equation}   
where $p_{\rm e}(Z;t)$ denotes the concentration PDF in 
the enriched/polluted part of the flow, and it satisfies that $\lim_{Z\to 0} \int_0^Z p_{\rm e} (Z';t) dZ' =0$. 
The Laplace transform of eq.\ (\ref{decomp}) gives, 
\begin{equation}
\hat{p}(\zeta; t) = P(t) + \hat{p}_{\rm e}(\zeta; t), 
\label{lapdecomp}
\end{equation}
where $\hat{p}_{\rm e}(\zeta; t)$ is the Laplace transform of $p_{\rm e}(Z;t)$. 
In the limit $\zeta \to +\infty$, $\hat{p}_{\rm e}(\zeta; t)$ approaches zero 
because $\lim_{Z\to 0} \int_0^Z p'_{\rm e}(Z';t) dZ' =0$. 

Inserting eq.\ (\ref{lapdecomp}) to  eq.\ (\ref{cconvolution}) for 
the model of Venaille and Sommeria (2007), and taking the 
limit $\zeta \to +\infty$, we find, 
\begin{equation}
\frac{d P(t)}{dt} =  \frac{P \ln(P)}{\tau_{\rm con}},
\label{pfcconv}
\end{equation}
where $\tau_{\rm con} \equiv \gamma^{-1}$, and we used the fact 
that $\hat{p}_{\rm e}(\zeta; t) \to 0$ and $\zeta \partial_\zeta \hat{p}_{\rm e} (\zeta; t) \to 0$ as 
$\zeta$ approaches infinity. If $\tau_{\rm con}$ is constant, the equation 
is solved by,  
\begin{equation}
P(t)  = P_0^ {\exp(t/\tau_{\rm con})},
\label{pfcconvsolution} 
\end{equation}
which can also be obtained from the 
solution for $\hat{p}(\zeta; t)$ given in Venaille and Sommeria (2007). 
We will show that eq.\ (\ref{pfcconvsolution}) provides a useful fitting function for our simulation data for 
$P(Z_{\rm c}, t)$ with finite $Z_{\rm c}$ in a transonic flow.  

Similarly, we can derive an equation for the pristine fraction from the 
generalized version, eq.\ (\ref{nconvolution}), of the self-convolution models,  
\begin{equation}
\frac{dP}{dt} = -\frac{n}{\tau_{\rm con}} P(1-P^{1/n}).  
\label{pfnconv}
\end{equation}
Assuming both $n$  and $\tau_{\rm con}$ are constant with time,  the solution of the equation is,  
\begin{equation}
P(t) = \frac{P_0}{\left[P_0^{1/n} + (1-P_0^{1/n} ) \exp\left( t /\tau_{\rm con} \right)  \right]^n}.  
\label{pfnconvsolution}
\end{equation}
For $n=1$, the equation reduces to eq.\ (\ref {pfintegralsolution}) for the nonlinear 
integral models, and in the limit of $n \to \infty$, it approaches eq.\ (\ref{pfcconvsolution}) for 
the continuous convolution model. We will use eq.(\ref{pfnconvsolution}) to fit our
simulation results for scalars in a highly supersonic flow, taking $\tau_{\rm con}$ and $n$ as fitting parameters. 



\section{Numerical Simulations}

To test the theoretical models and fix their parameters, 
we carried out numerical simulations for mixing 
in hydrodynamic turbulent flows using the FLASH 
code (version 3.2), a multidimensional hydrodynamic code 
(Fryxell et al.\  2000) that solves the Riemann problem 
on a 
Cartesian grid using a directionally-split Piecewise-Parabolic 
Method (PPM) solver (Colella \& Woodward 1984; 
Colella \& Glaz 1985; Fryxell, M\" uller, \& Arnett 1989).  
We evolved the hydrodynamic equations, 
\begin{gather}
\frac{\partial \rho}{\partial t} + {\bf \nabla} \cdot (\rho {\bf v}) = 0,  \notag \\
\frac{\partial {\bf v}} {\partial t} +  {\bf v} \cdot {\bf \nabla v} = -\frac{ {\bf \nabla} p}{\rho} +{\bf f},
\end{gather}
on a 512$^3$ grid for a domain of unit size with periodic boundary conditions. 
We adopted an isothermal equation of state, $p=\rho C_{\rm s}$, with a 
constant sound speed, $C_{\rm s}$. The isothermal equation of state is 
commonly used to imitate the nearly constant temperature 
in some interstellar environments, and is a convenient 
assumption to investigate the effects of compressibility 
in interstellar turbulence. Our code does not explicitly incorporate a viscosity term, and the kinetic 
energy is dissipated by numerical diffusion. A large-scale solenoidal external 
force, $\bf{f}$, was applied to drive and maintain the turbulent flows. This driving force was taken to be a 
Gaussian stochastic vector with an exponential temporal correlation function. 
We  generated  ${\bf f}$ in Fourier space and included all independent modes 
with wave numbers in the range from $2\pi$ and $6\pi$. Each independent mode 
was given the same amount of power. We defined a forcing length scale 
as $L_{\rm f} \equiv \int  \frac{2 \pi} {k} \mathcal{P}_{\rm f}(k)d{\bf k}/\int \mathcal{P}_{\rm f}(k)d{\bf k}$, 
with $ \mathcal{P}_{\rm f}(k)$ being the power spectrum of the driving force, 
and found  that  $L_{\rm f}$ was equal to 0.46 box size for our driving scheme.

We adjusted the amplitude of the driving force to obtain a 
transonic flow with rms Mach number $M=0.9$ and a 
supersonic flow with $M=6.2$. We refer to the two flows as flow 
A and flow B. The rms Mach number was defined as the 
density-weighted rms velocity, $v_{\rm rms}$, divided by 
the sound speed, $C_{\rm s}$, and was computed from the temporal 
average after the flow reached a statistical steady 
state. We defined a flow dynamical timescale as $\tau_{\rm dyn}\equiv L_{\rm f}/v_{\rm rms}$. 
The simulation setup for the turbulent flows is 
the same as that in Pan \& Scannapieco (2010), to which we refer the interested 
reader for details.  

To study mixing, we solved the advection equation for a number 
of decaying scalars, which were added to the flow once the turbulence 
had become fully developed and statistically stationary. The initial 
concentration field of the decaying scalars was  bimodal, consisting of pure 
pollutants and completely unpolluted flow. The initial pollutant 
region was chosen to be a single cube located right at 
the center of the simulation box. Within this cube, we set the 
concentration field, $C$ to be unity, i.e., the flow material there 
was taken be pure pollutants, and outside of the cube 
we set $C=0$, i.e., the flow there was completely pollutant 
free. This initial condition was chosen for its simplicity, and it 
suffices for the purpose of illustrating the general problem 
and testing the theoretical models. 

An important parameter for the initial condition is the pollutant fraction, 
$P_1$,  i.e., the ratio of the pollutant mass to the total mass in the 
simulation box. Clearly, the fraction fixes the initial pristine fraction, $P_0 =1-P_1$. 
We considered three scalars in each 
of our flows and set the initial pollutant fraction to be $P_1=$ 0.5, 0.1 and 0.01, 
respectively. In the $M=0.9$  flow, we name the three scalars 
with $P_1$= 0.5, 0.1, and 0.01 as  A1, A2 and A3, 
respectively. The corresponding cases in the $M=6.2$ flow 
are named B1, B2 and B3. The exact values for $P_1$ were 
achieved by tuning the size of the pollutant regions. Smaller values of $P_1$ 
would be also of interest for mixing of heavy elements in the 
interstellar media of early galaxies. However, 
for $P_1 \ll 0.01$, the size of the pollutant region becomes smaller than the integral 
scale of our simulated flows. This gives rise to complications in the 
evolution of the unpolluted (or slightly polluted) fraction. Smaller 
initial pollutant fractions will be investigated in a followup study.  

Similar to the case of kinetic energy dissipation, the scalar 
dissipation (or homogenization) is also through numerical 
diffusion in our simulations. The diffusion scale is thus close to the resolution 
scale. To examine whether our results depend on the amplitude of numerical 
diffusion, we performed the same runs at a lower resolution, $256^3$,  
and conducted a convergence study. We found that the 
timescale for the evolution of $P(Z_{\rm c}, t)$ with 
$Z_{\rm c} \sim 10^{-8}$ already converged at the resolution $512^3$.      

\section{Results}

\subsection{The Concentration Field}
\begin{figure}
\includegraphics[height=2.35in]{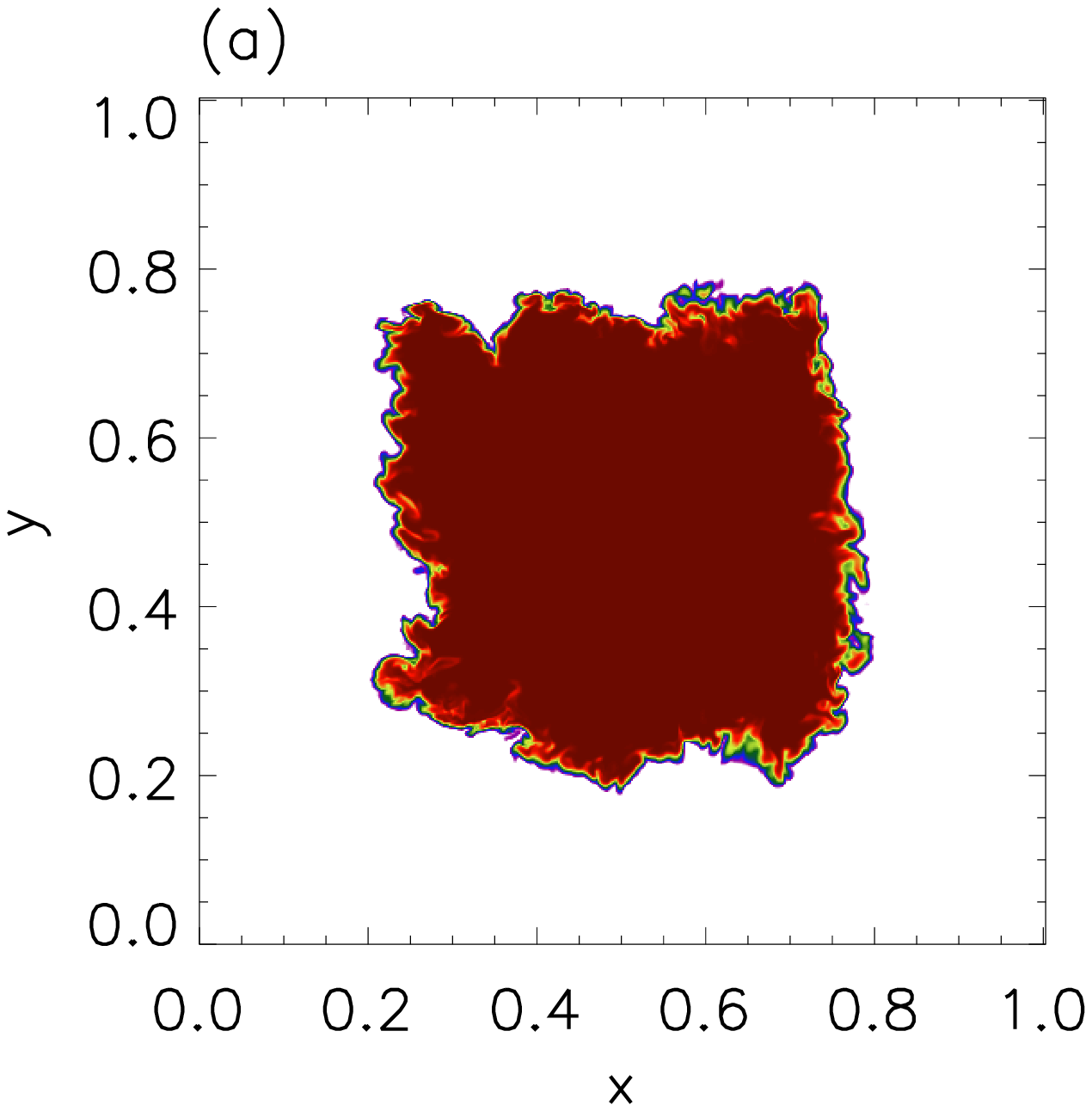}
\includegraphics[height=2.35in]{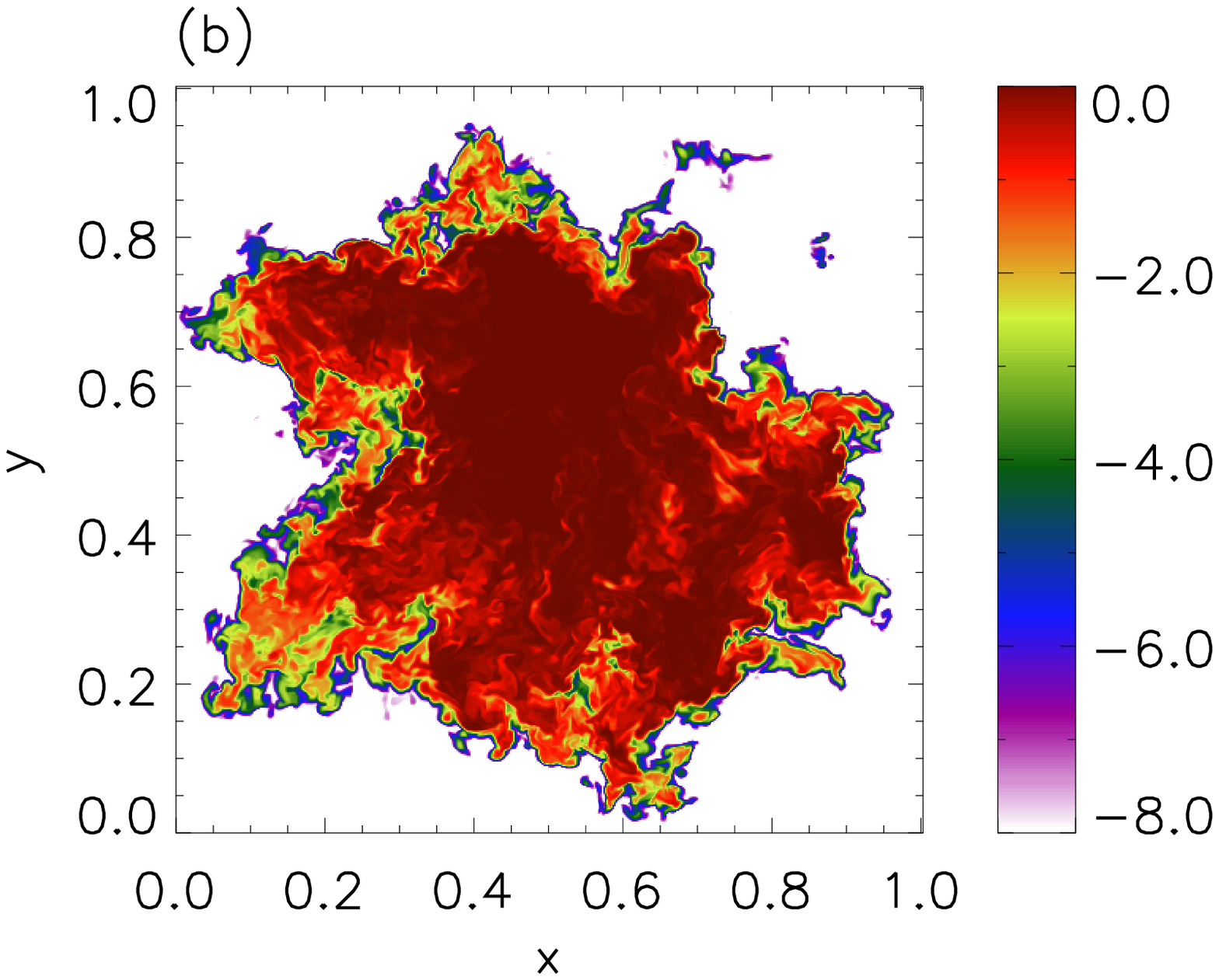}
\includegraphics[height=2.35in]{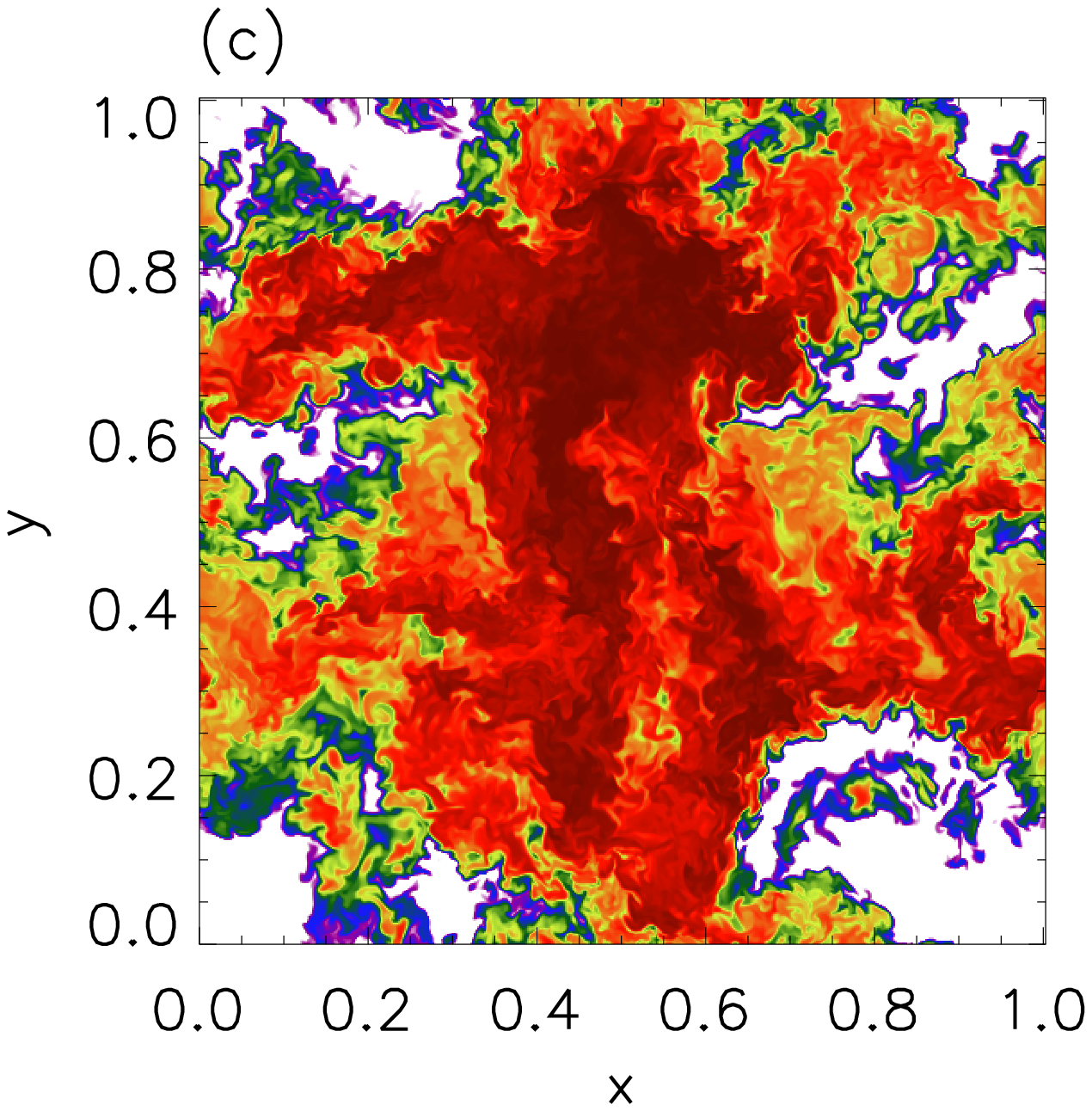}
\includegraphics[height=2.35in]{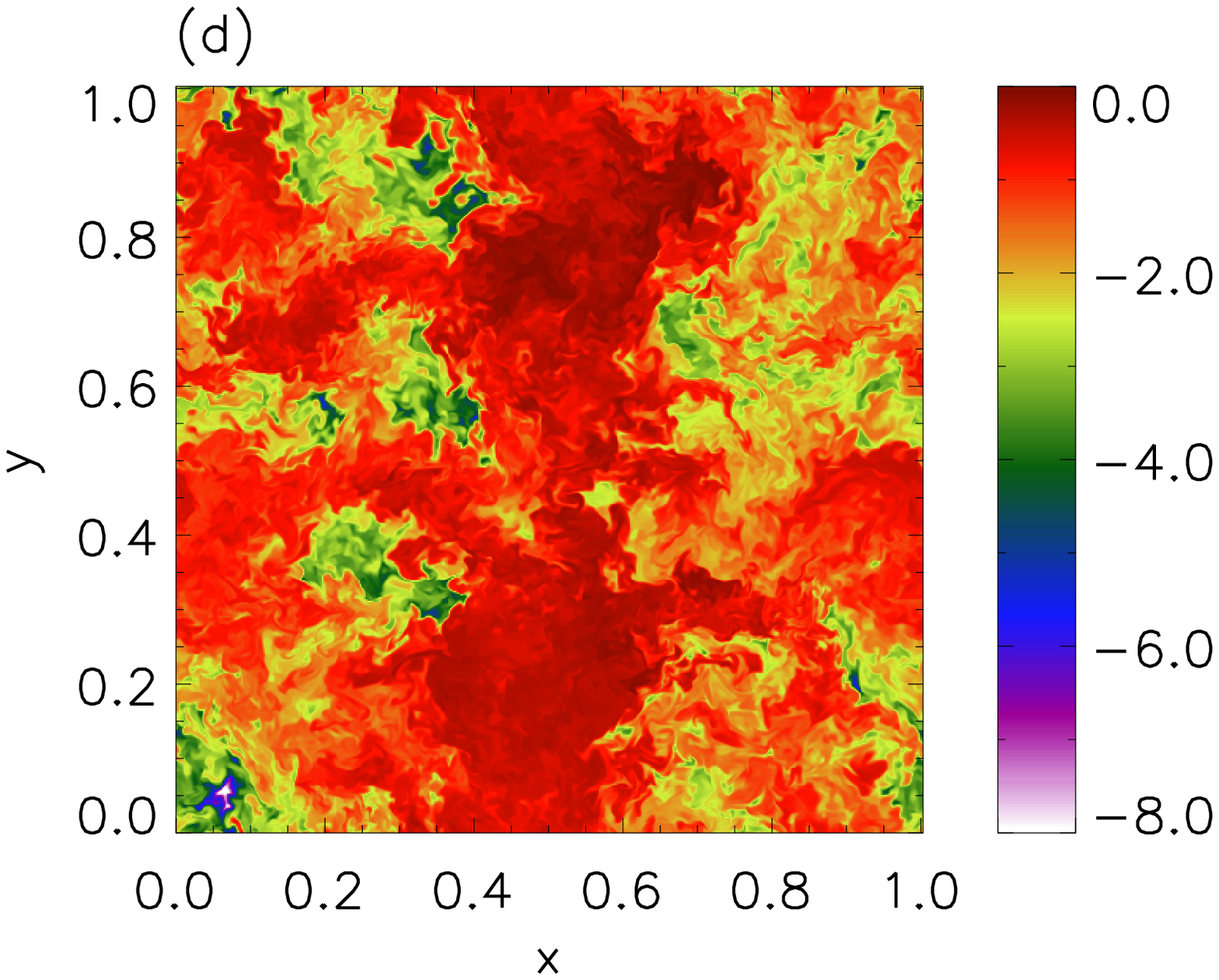}
\caption{Log of the concentration field of scalar A2 
on a slice of the simulation grid at snapshots  
with $t=$ 0.12 (a), 0.5 (b), 0.9 (c) and 1.5 $\tau_{\rm dyn}$ (d). 
The scalar is advected in the $M=0.9$ flow. 
The size of the initial pollutant cube is 0.47 box size.
The color table ranges from $10^{-8}$ to 1, 
with the white color representing 
regions with concentration $Z \le 10^{-8}$. 
}
\label{fig:image0.9}
\end{figure}

\begin{figure}
\includegraphics[height=2.35in]{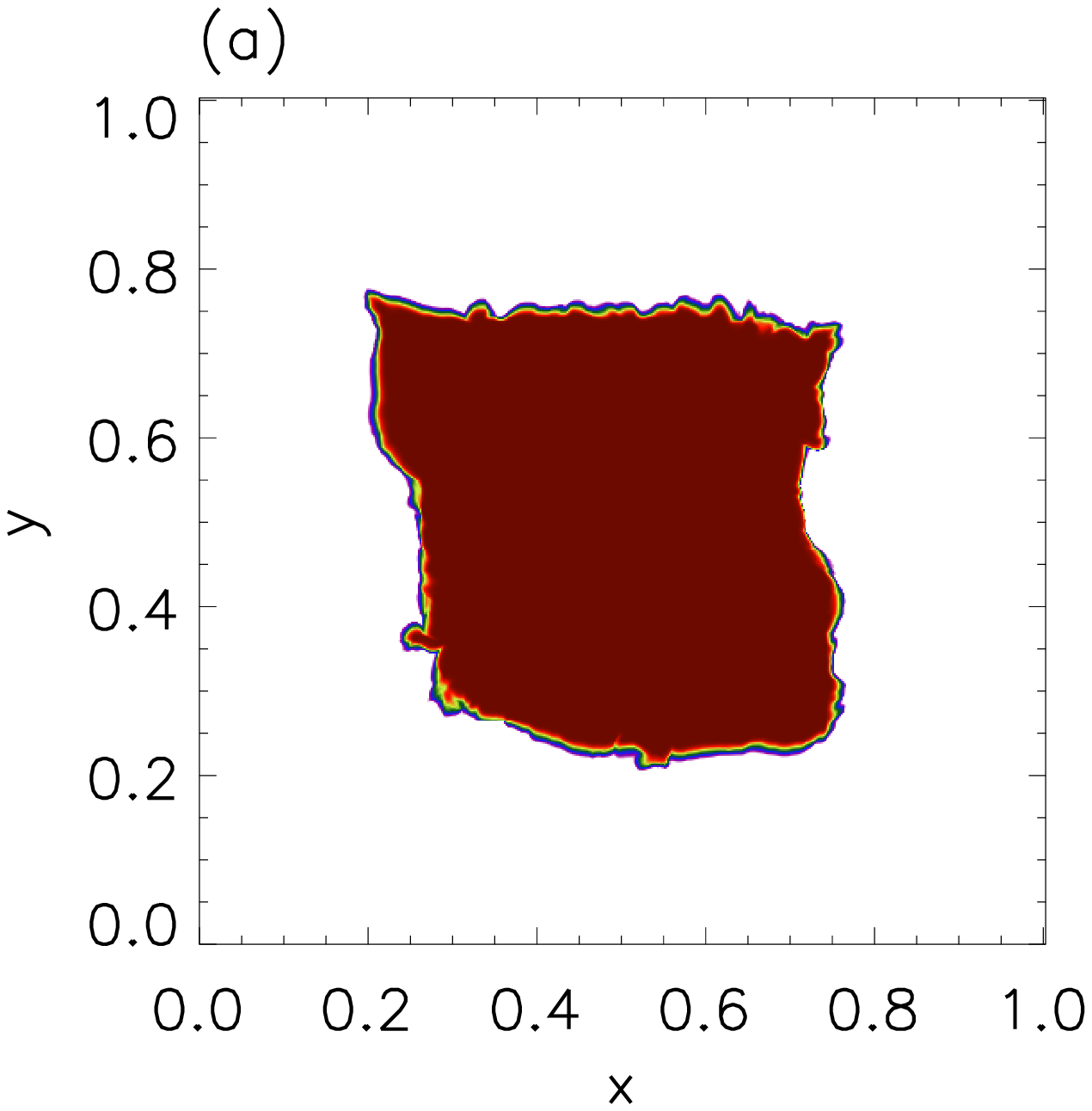}
\includegraphics[height=2.35in]{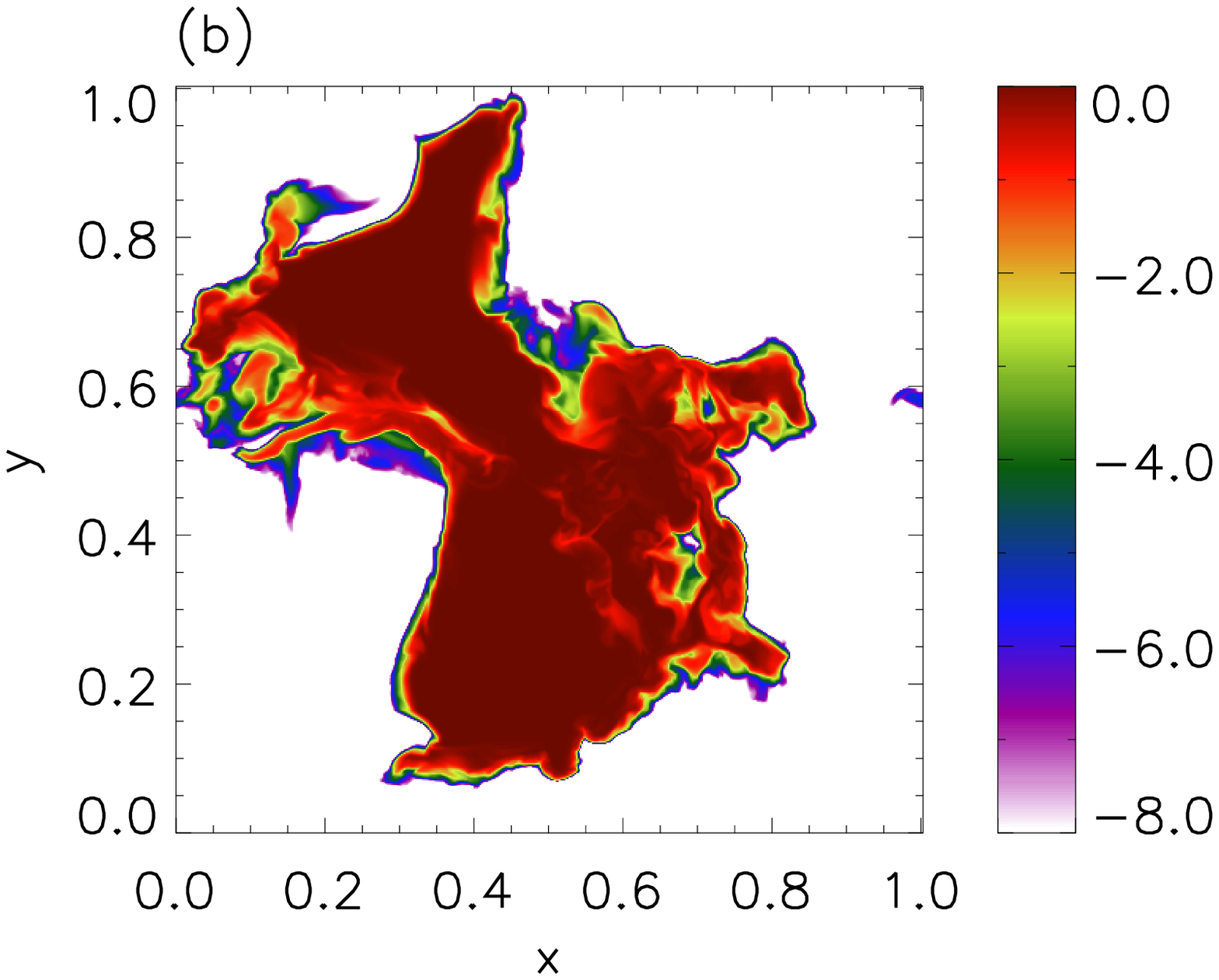}
\includegraphics[height=2.35in]{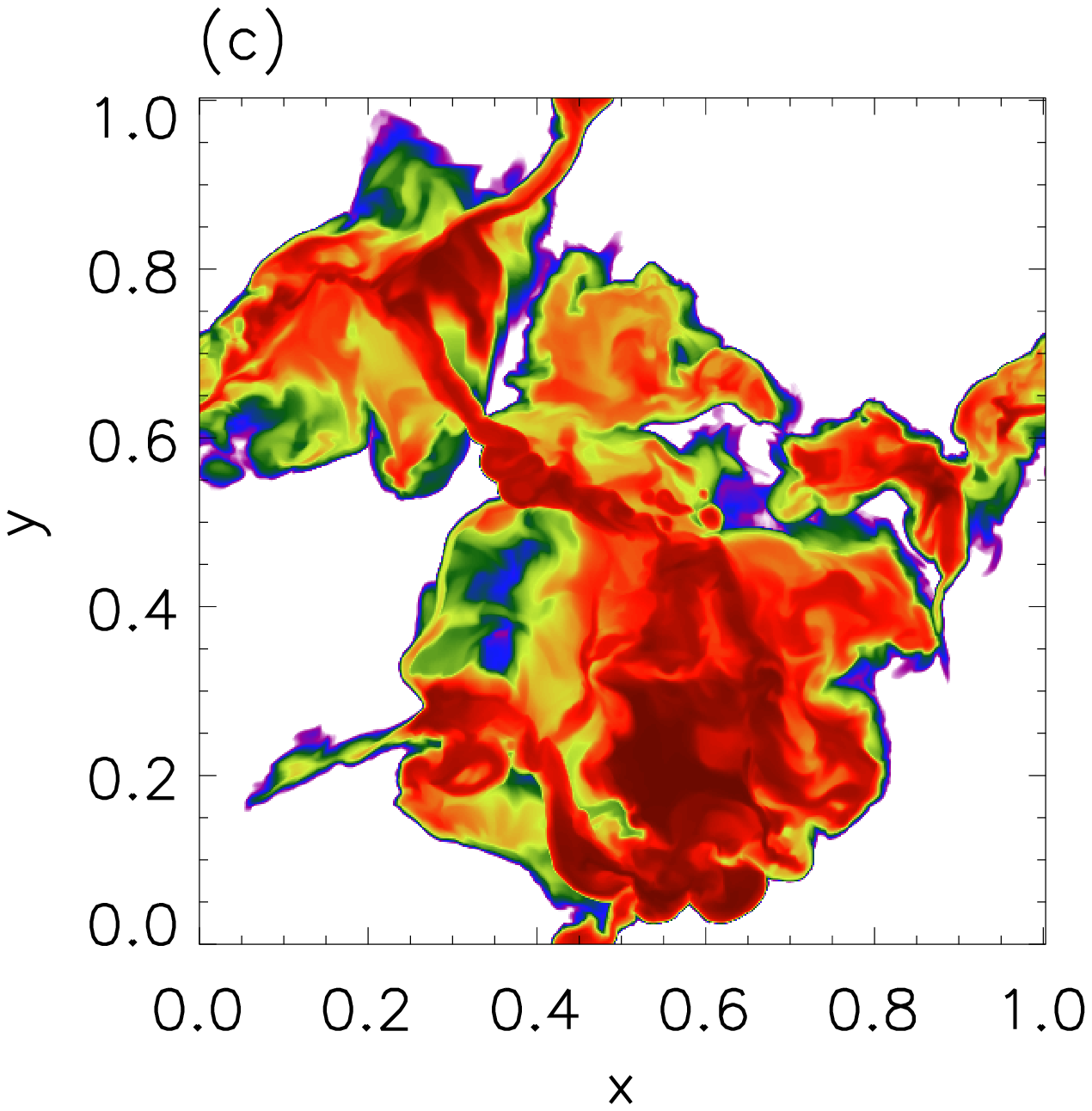}
\includegraphics[height=2.35in]{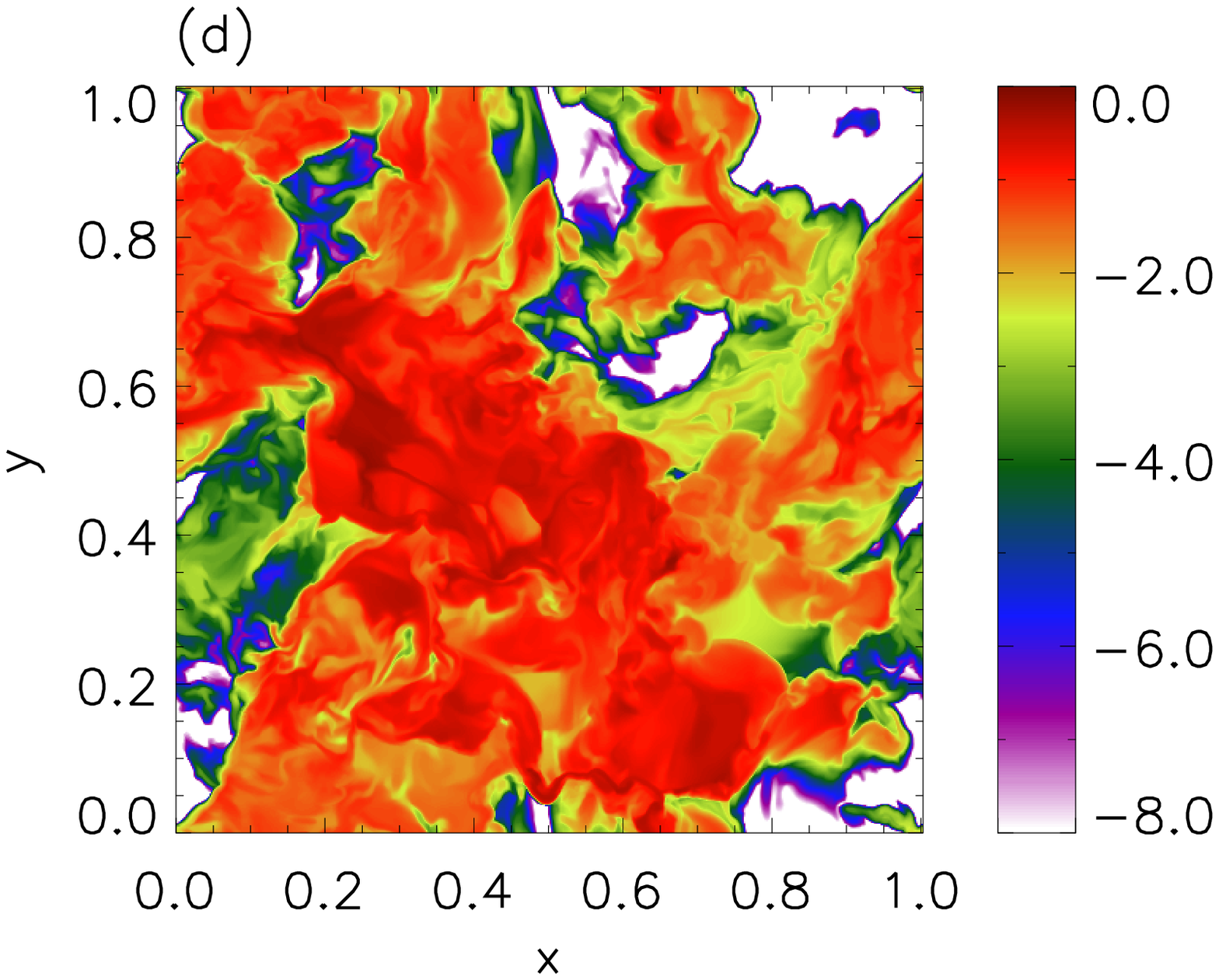}
\caption{Same as Fig.\ \ref{fig:image0.9}, but for scalar B2 
in the  $M=6.2$ flow. The four snapshots, (a), (b), (c) and (d), 
correspond to $t=$ 0.11, 0.65, 1.1 and 1.7 $\tau_{\rm dyn}$, 
respectively. The size of the initial pollutant cube is also 
0.47 box size, and the initial pollutant fraction is 0.1. }
\label{fig:image6.2}
\end{figure}

In Fig.\ \ref{fig:image0.9}, we plot the evolution of the 
concentration field of scalar A2 in our simulated flow 
with $M=0.9$. The four panels correspond to the 
log of the concentration field on a slice ($z=0.5$) of the 
simulation grid at four snapshots with $t=$0.12, 0.5, 0.9 and 
1.5 $\tau_{\rm dyn}$, respectively. The color table is in logarithmic 
scale and the lower limit was chosen to be 
10$^{-8}$, so that the part of the flow with concentration 
below a small threshold, $Z_{\rm c}$, is visible in 
the figure. The size of initial pollutant 
at the center of the box was set to be 0.47 in units of the box size, 
such that the initial pollutant fraction is 0.1. With time, the turbulent flow 
transports and spreads out the pollutants, exposing them to more and 
more pristine fluid elements. Turbulent stretching by vortices and shear 
continuously produces concentration structures at 
smaller and smaller scales.
Pristine fluid elements are contaminated when encountering 
a pollutant/polluted structure within a distance smaller 
than the diffusion scale.  At $t= 1.5 \tau_{\rm dyn}$, 
almost the entire flow is polluted, and the mass fraction 
of the flow with $Z \le 10^{-8}$ becomes negligibly small. Cliff structures 
typical of passive scalar fields advected in incompressible 
turbulence are clearly observed in panels (c) and (d).

Fig.\ \ref{fig:image6.2}  shows the concentration field of 
scalar B2 in our $M=6.2$ flow. At the four snapshots 
selected here, the density-weighted concentration 
variances are close to those for scalar A2 at the 
corresponding snapshots in Fig.\ \ref{fig:image0.9}. 
It appears that, at similar variances, the unpolluted 
volume in the $M=6.2$ flow is significantly larger 
than in the $M=0.9$ case. The existence of strong 
compressions and expansions in a highly supersonic 
flow gives rise to a very different geometry for the 
scalar field. Most of the volume in a highly compressible 
flow is occupied by expanding events, and the flow 
expansion tends to produce more coherent scalar 
structures, as a passive scalar follows the expansion. 
Therefore, scalar B2 appears to be smoother than 
A2 in Fig.\ \ref{fig:image0.9}. The edge-like structures 
in Fig.\ \ref{fig:image6.2} are produced by the 
compression of velocity shocks, which amplifies the 
scalar gradient across the shocks. Although the 
visual impression of Fig.\ \ref{fig:image6.2} is 
dominated by the effect of compressible modes, it 
is actually the solenoidal modes that contain the 
majority of kinetic energy in the flow and provide the 
primary contribution to the scalar cascade even at 
high Mach numbers (Pan \& Scannapieco 2010). 
The interested reader is referred to Pan and 
Scannapieco (2010, 2011) for detailed discussions of 
scalar structures as a function of the flow Mach 
number and the relative role of solenoidal and 
potential modes for mixing in supersonic turbulence.  



\subsection{The PDF Evolution}

\begin{figure}
\centerline{\includegraphics[width=1\textwidth]{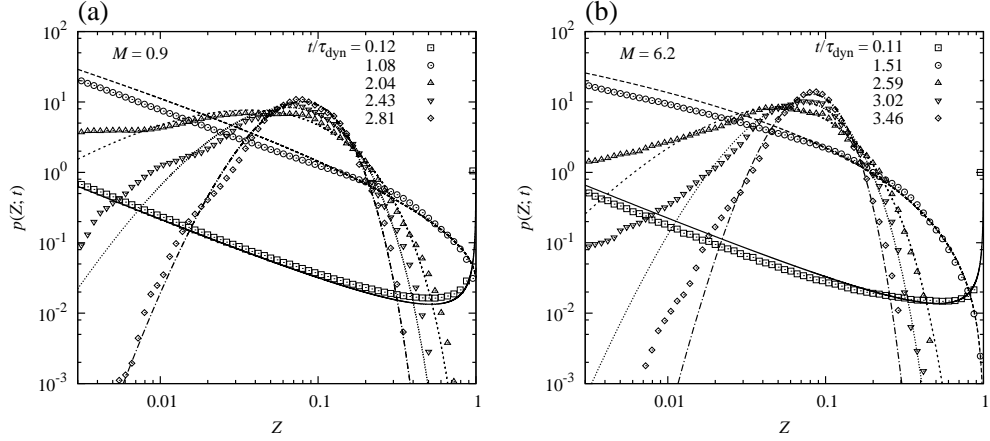}}
\caption{PDF evolution of scalars A2 (a) and 
B2 (b) in the $M=0.9$ and $M=6.2$ flows, respectively.  
The initial pollutant fraction, $P_1$, for the two scalars is 0.1. 
Lines are the predicted PDFs from the mapping closure 
model with the same values of variance as in the simulations. 
}
\label{fig:pdf}
\end{figure}

Fig.\ \ref{fig:pdf} plots the PDFs of scalars A2 (left panel) and B2 
(right panel) as a function of time. The two scalars were evolved 
in our simulated flows with $M=0.9$ 
and $M=6.2$, respectively. The initial pollutant fraction, $P_1$, 
is 0.1 for the two scalars, meaning that the amplitude of the 
initial spike at $Z=0$ is 9 times higher than that at $Z=1$. 
Turbulent mixing reduces the heights of the two 
spikes, and gradually fills the concentration space in 
between. Eventually a central peak forms around the 
average concentration, and then the PDF narrows down 
toward the peak. 

The lines in Fig.\ \ref{fig:pdf} are the prediction of the mapping 
closure model.  At each time, the predicted PDF has the 
same value of variance as that from the simulation (data points). 
This is equivalent to properly choosing the timescale 
${\lambda^2_\theta(t)}/\kappa$ in eq.\ (\ref{mappingeq})
so that the variance evolution from the model matches the 
simulation result. The model prediction is in good agreement 
with the data at the central part and the right (high-$Z$) tail 
of  the scalar PDF. However, the mapping closure considerably 
underestimates the left PDF tail at intermediate to late 
times. A detailed discussion of the discrepancy between the 
prediction of the Gaussian mapping closure and simulation 
results at late times is given in Girimaji (1992). The weakness of the 
mapping closure is also discussed by Duplat and Villermaux (2008). 
It appears that, for scalar A2, the agreement between the mapping 
closure and the simulation data becomes better in the long time limit. 


In the right panel of Fig.\ \ref{fig:pdf} for scalar B2 in the $M=6.2$ 
flow, we see that at early times there is also an acceptable 
agreement between the mapping closure prediction and the 
simulation results. At later times, the discrepancy at 
the left tails between the model and the data is larger than 
the $M=0.9$ case. This is because the PDF tails 
broaden with increasing Mach number, as previously found in the 
simulations of Pan \& Scannapieco (2010). The origin of 
this effect was argued to be related to the increasing 
degree of intermittency of the velocity field as the Mach 
number increases.  

The mass fraction of unpolluted or slightly polluted flow with 
$Z \lsim 10^{-8} -10^{-5}$ corresponds to the far left tail of the 
PDF with $Z$ well below the minimum value shown in 
Fig.\ \ref{fig:pdf}. Therefore, Fig.\ \ref{fig:pdf} does not 
contain direct information for the part of the PDF of our 
primary interest. Nevertheless, the left PDF tails shown 
in Fig.\ \ref{fig:pdf} imply that the mapping closure model 
is likely to significantly underestimate the unpolluted fraction 
at intermediate to late times especially for scalar B2. 
The expectation is confirmed in \S 6.4.  

\begin{figure}
\centerline{\includegraphics[width=1\textwidth]{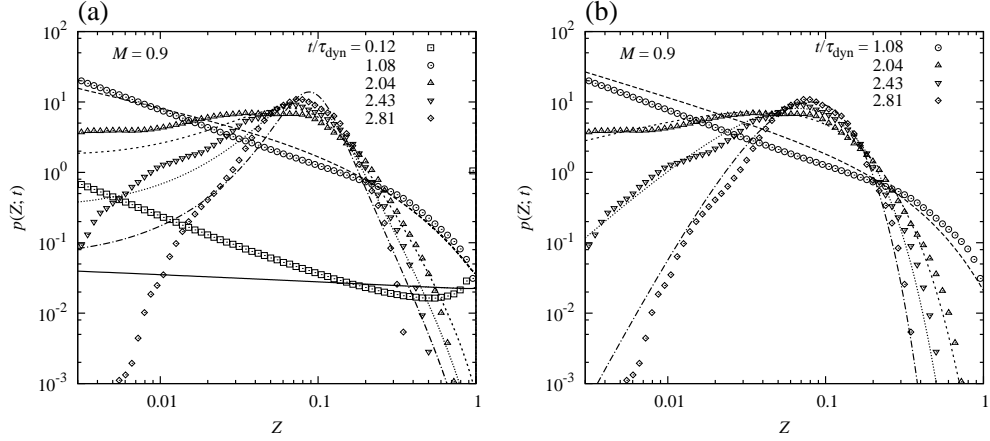}}
\caption{PDF evolution of scalar A2  in the $M=0.9$ 
flow. Lines are predictions of two models. 
(a): the nonlinear integral model with uniform $J(Z; Z_1, Z_2)$ (eq.\ (\ref{dopjan})). 
(b): Gamma distributions as predicted Villermaux and Duplat (2003).}  
\label{fig:pdf2}
\end{figure}

In the left panel of Fig.\ \ref{fig:pdf2}, we compare 
the prediction of the nonlinear integral model with 
uniform $J(Z; Z_1, Z_1)$ to the simulation data for 
scalar A2 in the $M=0.9$ flow. The performance of 
this model for the PDF evolution is poor. At very 
early times, the predicted PDF appears to be flat in 
between the initial spikes, reflecting a``memory" of 
the uniform function $J(Z; Z_1, Z_1)$.  As mentioned 
earlier, a problem of the nonlinear integral models is 
that they predict excessively fat tails in the long time 
limit. This is seen in the left panel of Fig.\ \ref{fig:pdf2}, 
which shows that at large $t$ the model significantly 
overestimates the left tails. We find that at late times 
the nonlinear integral model also overestimates the 
PDF tails for scalars in the $M=6.2$ flow (not shown). 
Although the nonlinear integral models do not give 
good predictions for the PDF evolution, we find that 
they provide useful fits to the pristine fraction in 
certain physical regimes (see \S 6.4).  

In the right panel of Fig.\ \ref{fig:pdf2}, we fit the PDF of 
scalar A2 in the $M=0.9$ flow with the Gamma distribution, 
eq.\ (\ref{gamma}), predicted by the model of Villermaux 
and Duplat (2003). For each line, the value of $n$ is 
chosen such that the variance of the Gamma distribution 
is equal to that from the simulation. The scalar PDF at 
$t \lsim 1 \tau_{\rm dym}$ does not have a Gamma 
distribution shape, and is not 
shown in this panel. At the four selected times in the figure, 
however, the Gamma distributions fit the simulation data 
quite well. The agreement is significantly better than the 
mapping closure for $t$ between $1.8$ and 2.6 $\tau_{\rm dyn}$ 
(corresponding to $1.1 \le n \le  4$). We find that $n(t)$ 
increases exponentially with time, corresponding to 
the exponential decay of the scalar variance at late times 
(see \S 6.3), as the variance of the Gamma distribution, 
eq.\ (\ref{gamma}), goes like $\langle Z \rangle^2/n$. 
This is in contrast to the experimental result, 
$n(t) \propto t^{5/2}$, found by Villermaux and Duplat (2003). 
The reason for the difference is that our simulated flows 
are maintained at a steady state by a driving force and are 
statistically homogeneous, while the experiments  by 
Villermaux and Duplat (2003) are for decaying flows 
dominated by a mean shear. An exponential decay is 
also found in a sustained flow by Villermaux et al.\ (2008). 

We point out that the continuous convolution model of Venaille 
and Sommeria (2007) predicts that, if the scalar PDF is given 
by a Gamma distribution at a given time, then the PDF will 
remain a Gamma distribution at all subsequent times. 
Therefore, if one starts to use the continuous convolution model at a 
time when the scalar PDF has evolved to a Gamma distribution, its 
prediction for later times would be the same as the model of 
Villermaux and Duplat (2003). However, unlike Villermaux 
and Duplat (2003), the model of Venaille and Sommeria 
(2007) does not predict that the Gamma distribution is 
an attractive solution that the scalar PDF always reaches at 
the late evolution stage.  

For scalar A2 in the $M=0.9$ flow, the model of Villermaux and 
Duplat (2003) starts to apply at $t= 1-2 \tau_{\rm dyn}$.  At this 
time, the pristine mass fraction has already decreased to 
very small values, and thus the model is not suitable to study the pristine fraction. 
We also tried to fit Gamma distributions to the scalar PDFs in 
the $M=6.2$ flow, and found they underestimate the left tails, 
which are broader than the $M=0.9$ case.     


In summary, we found that in the $M=0.9$ flow the mapping 
closure gives acceptable fits to the scalar PDF  at early times, 
but significantly underestimates the left PDF tails at late times. 
Starting from $\simeq 1.8 \tau_{\rm dyn}$, the scalar PDF is better fit by a 
Gamma distribution, which is predicted by the model of Villermaux and 
Duplat (2003). Since all the models we considered were originally developed 
to study mixing in incompressible turbulence, they were not expected to 
perform well in highly compressible flows. In the $M= 6.2$ flow, 
the left PDF tail is broader, and no models were found to give satisfactory 
predictions for the scalar PDF at late times. 


\subsection{The Variance Decay}

\begin{figure}
\includegraphics[width=1\textwidth]{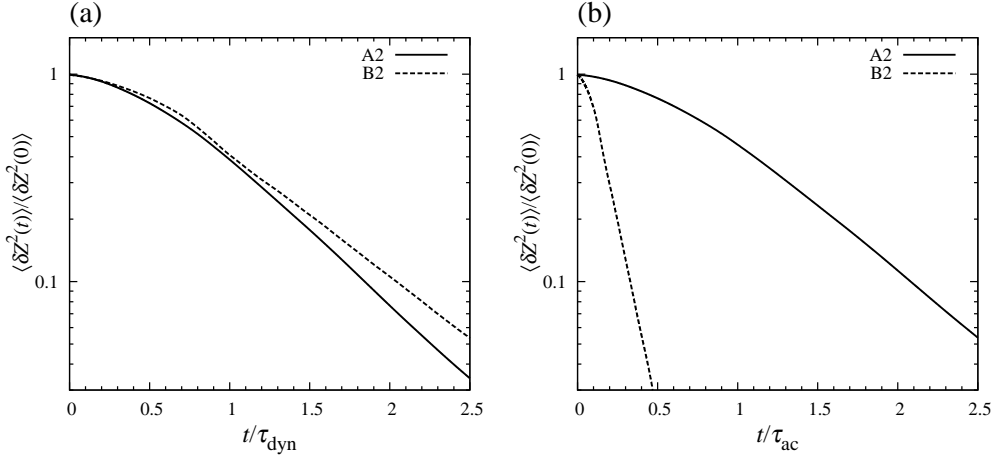}
\caption{Evolution of the density-weighted 
concentration variance for scalars  A2 and B2 
in $M=0.9$ and $M=6.2$ flows, respectively. The variance is normalized 
to its initial value. (a): time normalized to the flow dynamical timescale. 
(b): time normalized to the acoustic timescale.}
\label{fig:var}
\end{figure}

In Fig.\ \ref{fig:var}, we show the evolution of the density-weighted 
concentration variance for scalars A2 and B2 in the $M=0.9$ and 
$M=6.2$ flows, respectively. In the left panel, the time is normalized to the 
flow dynamical time $\tau_{\rm dyn}$, while in the right panel 
it is normalized to the acoustic time $\tau_{\rm ac}$ defined as $L_{\rm f}/C_{\rm s}$,  
the time for the sound wave to cross the driving scale of the 
flow, $L_{\rm f}$. From the definition, we have $\tau_{\rm ac} = M \tau_{\rm dyn}$. 
Therefore, the variance decay curves shift to the right  
by a factor of 1.1 for scalars A2 and to 
the left by a factor of 6.2 for B2, when the normalization 
timescale is switched from $\tau_{\rm dyn}$ to $\tau_{\rm ac}$.  
From Fig.\ \ref{fig:var}, it is clear that the flow dynamical time 
is a more relevant timescale for the scalar variance decay in 
supersonic isothermal turbulent flows.   

As seen in the left panel of Fig.\ \ref{fig:var}, at early times 
the variance decreases slowly, corresponding to the initial 
development of scalar structures toward small scales. 
In this transient period, the parameter $\gamma(t)$ in 
the models presented in \S 3.3 and \S 3.4 would be time-dependent 
if the model is required to match the scalar variance decay.  At $t \gsim 0.5 \tau_{\rm dyn}$, 
the variance decays exponentially, and $\gamma(t)$ would be constant.  
When normalized to the dynamical timescale, the variance decay of 
scalar A2 is slightly faster than B2, and the decay timescale is measured to be $\tau_{\rm m} = 0.61$ 
and $0.73 \tau_{\rm dyn}$, for A2 and B2, respectively. These results are consistent 
with Pan \& Scannapieco (2010), who found that the variance decay timescale in units of the flow dynamical 
timescale has a weak dependence on the flow Mach number, increasing 
by $\lsim $20\% as $M$ goes from 1 to 6. This slight increase 
is due to the fact that compressible modes are less efficient at enhancing 
the mixing rate  than solenoidal modes (Pan \& Scannapieco 2010).  

We also measured $\tau_{\rm m}$ for the other four scalars included 
in our simulations, and found that, in each flow, the mixing 
timescales of the three scalars are close to each other. The timescale 
for the third scalar (i.e., A3 or B3) is slightly smaller than the other two. 
On average, the mixing timescale is $\tau_{\rm m}  \simeq 0.6 \tau_{\rm dyn}$ 
for the three scalars in the $M=0.9$ flow. In the $M=6.2$ flow, 
the average mixing timescale is $\simeq 0.7 \tau_{\rm dyn}$.

\subsection{The Fraction $P(Z_{\rm c}, t)$}

Next we computed the mass fraction, $P(Z_{\rm c}, t)$, of fluid 
elements with $Z$ smaller than different thresholds, $Z_{\rm c}$, 
from the simulation data. We found that the flow with exactly 
zero concentration (i.e., the special case with $Z_{\rm c}=0$) 
is erased rapidly by numerical diffusion. This is because in 
each time step the unpolluted computation cells adjacent 
to those with nonzero $Z$ obtain a finite (but tiny) 
concentration value due to numerical diffusion. 
Therefore, after a small number of time steps, no 
exactly pollutant-free cells were left in the simulation box. 
However, in such a short time, the degree of pollution in most cells 
due to numerical diffusion is extremely small, and the 
concentration level in these cells was much smaller than 
any threshold of practical interest. The rapid pollution 
of completely pristine mass by numerical diffusion in the 
simulations is similar to the effect of molecular diffusivity, 
which tends to reduce $P(t)$ to zero instantaneously, 
although the numerical diffusion probably has a 
different form and a much larger amplitude than the 
realistic molecular diffusivity.  
 
The threshold of interest for astrophysical applications is $Z_{\rm c} \gsim 10^{-8}$. 
For this finite threshold, the timescale at which $P(Z_{\rm c}, t)$ decreases is significant, 
and is on the order of the flow dynamical time. A comparison of the same 
simulation runs at two resolutions, 256$^3$ and $512^3$, shows 
that the timescale for $P(Z_{\rm c}, t)$ with $Z_{\rm c} \sim 10^{-8}$ is independent of the amplitude 
of numerical diffusion. These suggest that the reduction rate 
of $P(Z_{\rm c}, t)$ with $Z_{\rm c} \gsim 10^{-8}$ is mainly determined by 
the large-scale properties of the flow. The behavior of $P(Z_{\rm c}, t)$ as 
a function of time is similar for different values of $Z_{\rm c}$, 
given that $Z_{\rm c}$ is much smaller than 
the average concentration. The evolution timescale of $P(Z_{\rm c}, t)$ only has a weak logarithmic dependence 
on $Z_{\rm c}$. Below we only present results for $Z_{\rm c} = 10^{-8}$. 
Similar results are found for $Z_{\rm c}$ in the range from $10^{-8}$ to $10^{-5}$.    

\subsubsection{The $M=0.9$ Flow}

\begin{figure}
\centering\includegraphics[width=0.65\textwidth]{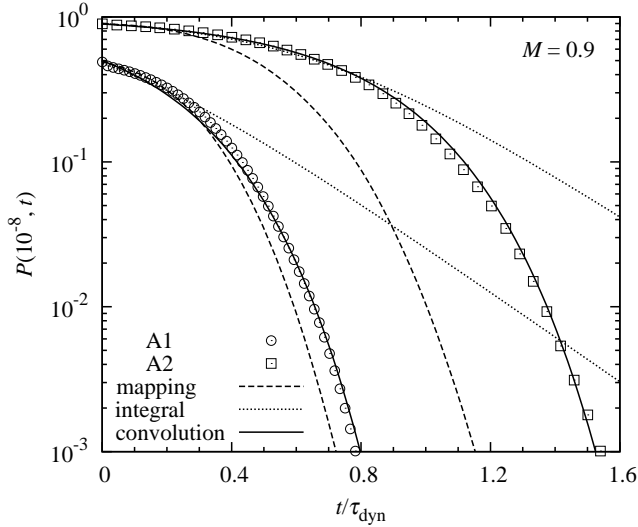} 
\caption{Mass fraction of fluid elements with 
$Z \le 10^{-8}$ for scalars A1 and A2 in the 
$M=0.9$ flow.  Dashed lines correspond 
to the prediction of the mapping closure model.   
Solids lines are the best fits using the continuous 
convolution model with $\tau_{\rm con} = 0.35$ and 
0.37 $\tau_{\rm dyn}$ for A1 and A2, respectively. 
The dotted lines are fits by eq.\ (\ref{pfintegralsolution}) 
from the nonlinear integral models.}
\label{fig:pfA1A2}
\end{figure}

In Fig.\ \ref{fig:pfA1A2}, we plot the mass fraction 
of fluid elements with $Z \le 10^{-8}$ as a function of 
time for scalars A1 and A2 in the $M=0.9$ flow.  
The initial pollutant fraction, $P_1$, is 0.5 and 0.1 
for the two scalars, respectively. The data points 
are results from the simulations. The dashed lines 
correspond to the prediction of the mapping closure model. For this model, the fraction, 
$P(Z_{\rm c}, t)$, at a given time is calculated from 
the predicted PDF with the same scalar variance as that 
in the simulation. In Fig.\  \ref{fig:pfA1A2}, we see the 
mapping closure model is in good agreement with the data points for scalar 
A1. However, the model prediction is well below 
the data points for scalar A2. This was expected from 
the left panel of Fig.\ \ref{fig:pdf}, which shows that the 
mapping closure model underestimates the left PDF tail 
of scalar A2 at intermediate to late times. We found that the 
model also underestimates $P(Z_{\rm c}, t)$ for scalar A3, 
which had an initial pollutant fraction of 0.01, 
and the discrepancy is even larger than the case of A2. 

The solid lines in Fig.\ \ref{fig:pfA1A2} are from the continuous 
convolution model of Venaille and Sommeria (2007), i.e., eq.\ (\ref{pfcconvsolution}), 
and they match the data points quite well. The timescales $\tau_{\rm con}$ 
used in the fits are $0.35$ and 0.37 $\tau_{\rm dym}$, respectively, 
for scalars A1 and A2. As discussed in \S 4.3, eq.\ (\ref{pfcconvsolution}) 
was originally derived for the fraction, $P(t)$, of exactly pollutant-free mass. 
The good agreement between eq.\ (\ref{pfcconvsolution}) 
and the simulation data shows that the model actually provides 
an excellent fitting function for the fraction, $P(Z_{\rm c}, t)$, 
with a finite (but small) threshold $Z_{\rm c}$. It also suggests 
that the continuous convolution process is a good physical 
description for the erasure of unpolluted (or slightly polluted) flow 
by turbulent mixing, if the initial pollutant fraction, $P_1$, is larger than $\sim 0.1$. 

The dotted lines in Fig.\ \ref{fig:pfA1A2} shows the fits using eq.\ (\ref{pfintegralsolution}) 
from the nonlinear integral model. The timescale $\tau_{\rm int}$ was chosen 
to be $0.28 \tau_{\rm dyn}$ and $0.3 \tau_{\rm dyn}$ for scalars A1 and A2, 
respectively. This model predicts an exponential decrease at late times, which 
is much slower than found in the simulation. The overestimate is 
because the model produces excessively broad PDF tails in the long time 
limit (see left panel of Fig.\ \ref{fig:pdf2}).  

\begin{figure}
\centering\includegraphics[width=0.65\textwidth]{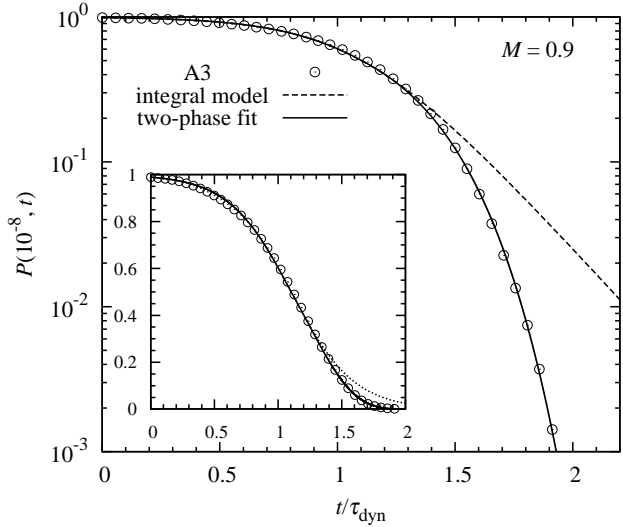} 
\caption{Mass fraction of fluid elements with 
$Z \le 10^{-8}$ for scalar A3 in the $M=0.9$ flow. 
The dashed line corresponds to the nonlinear 
integral model with $\tau_{\rm int} = 0.24 \tau_{\rm dyn}$.  
The solid line is the best fit obtained by combining 
the nonlinear integral model for the early phase 
and the continuous convolution model for 
the later phase. The timescales used in the 
two phases are $\tau_{\rm int} = 0.24 \tau_{\rm dyn}$ and 
$\tau_{\rm con} = 0.36 \tau_{\rm dyn}$, respectiely. 
The inset shows the same data points and 
lines, but with the vertical coordinate on a linear scale.}
\label{fig:pfA3}
\end{figure}

Fig.\ \ref{fig:pfA3} shows the evolution of $P(10^{-8}, t)$ 
for scalar A3, whose initial pollutant fraction $P_1= 0.01$. 
The inset plots the same data points and model 
fits, but the y-axis is on a linear scale. 
Unlike the case of scalars A1 and A2, the data 
points for scalar A3 cannot be well fit by the 
continuous convolution model with a single timescale 
right from the beginning. In fact, $P(10^{-8}, t)$ exhibits 
different behaviors at early and late times. 

The early phase can be well fit by eq.\ (\ref{pfintegralsolution}) 
from the nonlinear integral model with $\tau_{\rm int} = 0.24\tau_{\rm dyn}$. 
This is shown as a dashed line in Fig.\ \ref{fig:pfA3}, and from 
the inset we see the line matches the simulation data over 
an extended time range before $P(10^{-8}, t)$ decreases to 
$\simeq 0.3$. The timescale $\tau_{\rm int}$ 
used here corresponds to $\sim 0.4 \tau_{\rm m}$ 
since the variance decay timescale $\tau_{\rm m}$ for scalar 
A3 is about $0.6 \tau_{\rm dyn}$.  This is in between 1/2 
and 1/3 $\tau_{\rm m}$, the expected values of $\tau_{\rm int}$ 
for Curl's model and the model with uniform 
$J(Z; Z_1, Z_2)$, respectively (see \S 4.2). The dashed line 
starts to significantly overestimate the simulation results 
when $P(10^{-8}, t)$ becomes smaller than $\sim 0.3$.  
Again, this is because the nonlinear integral models significantly 
overpredict the PDF tails at late times.   

We find that  the late-time behavior of $P(10^{-8}, t)$ 
can be well described by the continuous convolution 
model. In fact, this model starts to give a satisfactory 
fit quite early, right after $P(10^{-8}, t)$ becomes smaller 
than $\sim 0.8$. The best-fit value of the timescale 
$\tau_{\rm con}$ is $0.36 \tau_{\rm dyn}$, which is almost 
the same as the values used to fit the data for scalars 
A1 and A2. This, together with the results for scalars A1 and A2, 
suggests that the continuous convolution model applies if the 
mass fraction of pollutants or the polluted flow is larger than 
$0.1-0.2$. We point out that there is an extended range of  
$P(10^{-8}, t)$ (from 0.3 to 0.8), where both the nonlinear 
integral model and the continuous convolution model 
can well match the data. 

We give a physical speculation for why the early phase 
of scalar A3 is better fit by the nonlinear integral model 
than the continuous convolution model. For this scalar, 
the amount of pollutants or polluted mass is small at 
early times. The limited availability of pollution sources 
leads to a relatively low frequency of contact between 
the polluted and unpolluted flow. As a consequence, the 
pollution process would involve less convolutions at this stage, 
and thus may be better captured by the nonlinear 
integral model, which can be roughly viewed as a 
discrete self-convolution model in Laplace space. 
The simulation results presented above suggest 
that the mixing events between the unpolluted and the 
polluted fluid elements become frequent 
enough to trigger the continuous convolution process, 
when the polluted fraction is larger than 0.1-0.2. 
 
The solid line in Fig.\ \ref{fig:pfA3} is obtained by 
combining the best fits for the early and late phases 
using the nonlinear integral model and the continuous 
convolution model, respectively. Clearly, this line is in excellent 
agreement with the simulation 
results. We connected the two phases at time, $t_{0.5}$, when $P(10^{-8}, t) = 0.5$, 
i.e., the second phase is fit by $0.5^{\exp[(t-t_{0.5})/\tau_{\rm con}]} $. 
As mentioned earlier, the best-fit timescales for the two phases are $\tau_{\rm int} = 0.24 \tau_{\rm dyn}$ and 
$\tau_{\rm con} = 0.36 \tau_{\rm dyn}$, respectively.   
The choice of connecting the two phases at 
$P(10^{-8}, t) = 0.5$ is somewhat arbitrary. In 
fact, combining the two models at any time with 
$P(10^{-8}, t)$ between $\simeq 0.8$ and $ \simeq 0.3$ 
yields satisfactory results. 

The timescales, $\tau_{\rm int}$ and $\tau_{\rm con}$, 
were set to be constant in all our fits to the simulation 
results for $P(Z_{\rm c},t)$. These timescales can 
be a function of time in general.  In fact,   from 
the consideration of the scalar variance decay, these 
timescales would be time-dependent at early 
times when the variance decay is slower than exponential, 
and then become constant at $t \gsim 0.5 \tau_{\rm dyn}$ 
when the exponential decay starts (see \S 6.3). 
The reason why assuming constant timescales 
applies perfectly for the evolution of the pristine mass 
fraction, but not for the scalar variance decay at all times 
is probably that the pollution of pristine flow 
is physically simpler. A fast variance decay relies on the full development 
of scalar structures toward the diffusion scale, and that is 
why the decay is slow at early times as the cascade just starts. 
On the other hand, the pollution of pristine mass does 
not need to wait for the scalar structures to be fully 
developed at small scales: the pollution occurs whenever the unpolluted fluid 
elements are brought into contact with the pollutants or the polluted 
flow. This happens right away once the pollutants are 
released into the flow. This is perhaps why using constant timescales, 
$\tau_{\rm int}$ and $\tau_{\rm con}$, right from the beginning 
could give successful fits to the evolution of $P(Z_{\rm c}, t)$.  
 

\subsubsection{The $M=6.2$ Flow}

\begin{figure}
\centering\includegraphics[width=0.65\textwidth]{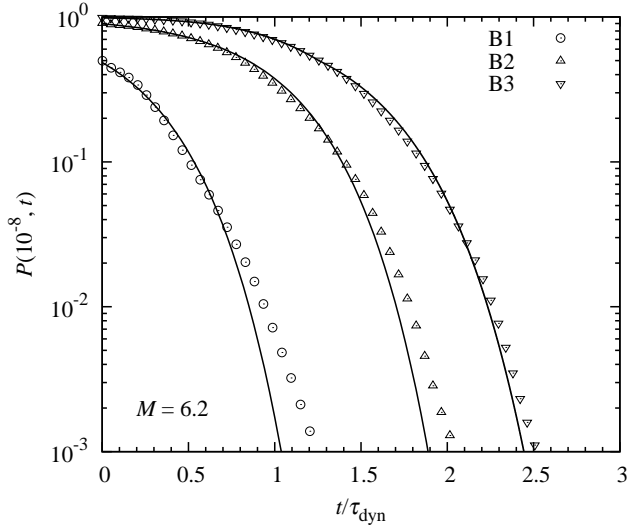} 
\caption{Mass fraction of fluid elements with 
$Z \le 10^{-8}$ for scalars B1, B2 and B3 in the 
$M=6.2$ flow. The solid lines for B1 and B2 
correspond to the fits by the continuous convolution 
model (eq.\ \ref{pfcconvsolution}) 
with $\tau_{\rm con} = 0.46\tau_{\rm dyn}$ for both cases. 
The line for B3 combines the nonlinear integral 
model with $\tau_{\rm int} =0.30\tau_{\rm dyn}$ for the 
early phase and the continuous convolution model with 
$\tau_{\rm con} = 0.51 \tau_{\rm dyn}$ for the later phase.
}
\label{fig:pfB1}
\end{figure}

Fig.\ \ref{fig:pfB1} shows our results for scalars B1, B2 and B3 
with initial pollution fractions of 0.5, 0.1, and 0.01, respectively, 
in the $M=6.2$ flow.  We find that the mapping closure model 
significantly underestimates $P(10^{-8}, t)$ for all the three 
scalars, and the discrepancy is much larger than the case 
of the $M=0.9$ flow. In Fig.\ \ref{fig:pfB1}, we attempt to fit the 
simulation data for scalars B1 and B2 with the continuous convolution model 
[eq.\ (\ref{pfcconvsolution})] of Venaille and Sommeria (2007), as 
in the $M=0.9$ case. The timescale, $\tau_{\rm con}$, used 
in the fitting lines is 0.46 $\tau_{\rm dyn}$ for both B1 and B2. Again 
the fitting curve for scalar B3 consists of two phases that connect at $P(Z_{\rm c}, t)=0.5$. 
The early phase is fit by the nonlinear integral model with $\tau_{\rm int}=0.3 \tau_{\rm dyn}$, 
and the later phase uses the continuous convolution model 
with $\tau_{\rm con} = 0.51\tau_{\rm dyn}$. The parameter 
choice here gives priority to satisfactorily matching the data points at early times. 

All the timescales chosen in Fig.\ \ref{fig:pfB1} are larger than the corresponding values used for 
scalars in the $M=0.9$ flow. This is caused by two 
effects. First, as mentioned earlier, when normalized to the flow 
dynamical time,  the variance decay timescale in the $M=6.2$ flow is 
slightly larger than in the $M=0.9$ case. Second, as shown in Fig.\ \ref{fig:pdf}, 
the left tail of the scalar PDF broadens with increasing Mach number 
of the advecting flow. This means that, with the same concentration 
variance, the scalar PDF in the $M=6.2$ flow contains a larger probability 
at low concentration levels. Both effects tend to result in a 
slower decrease of $P(Z_{\rm c}, t)$ in the flow with higher $M$. 
The second effect appears to be stronger than the first one, and 
it also explains why the same models that well match the results in 
the $M=0.9$ flow significantly underestimate 
$P(Z_{\rm c},t)$ in the $M=6.2$ flow at late times (see Fig.\ \ref{fig:pfB1}). 
The fitting quality of the lines in Fig.\ \ref{fig:pfB1} is good at early times 
when $P(Z_{\rm c},t) \gsim  0.1$, and generally acceptable 
before $P(Z_{\rm c},t)$ decreases to 0.01. Below that, however, significant discrepancy appears. 


\begin{figure}
\centering\includegraphics[width=0.65\textwidth]{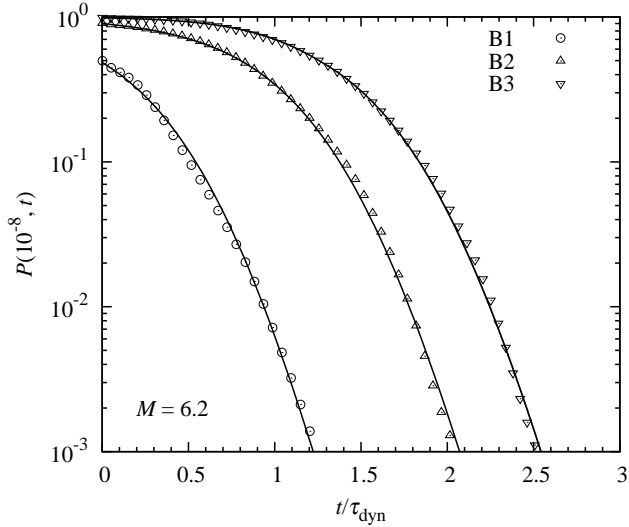} 
\caption{Mass fraction of fluid elements 
with $Z \le 10^{-8}$ for scalars B1, B2 and B3 
in the $M=6.2$ flow. The data points are the 
same as in Fig.\ \ref{fig:pfB1}. Eq.\ (\ref{pfnconvsolution})  
with $n=4.6$ is used to match the simulation data. 
The timescale, $\tau_{\rm con}$, in the equation 
is set to $0.40$ and $0.41 \tau_{\rm dyn}$ for scalars B1 and B2, 
respectively. The line for B3 is a combination 
of the nonlinear integral model with $\tau_{\rm int} = 0.30\tau_{\rm dyn}$ 
for the early phase and eq.\ (\ref{pfnconvsolution}) with $\tau_{\rm con} = 0.42 \tau_{\rm dyn}$ and $n=4.6$ 
for the later phase.
}
\label{fig:pfB2}
\end{figure}

We find that the simulation results for B1 and B2 in the $M=6.2$ flow 
can be better fit by eq.\ (\ref{pfnconvsolution}) from the generalized self-convolution 
model (see eq.\ (\ref{nconvolution}) in \S 3.4; Duplat \& Villermaux 2008). 
The parameter $n$ in this equation controls the shape of the fitting curve. Fig.\ \ref{fig:pfB2} 
shows our results using this equation to fit the simulation data. The data points here are the same as 
in Fig.\ \ref{fig:pfB1}. Eq.\ (\ref{pfnconvsolution}) with $n=4.6$
can well fit the simulation data for scalars B1 and B2 at all times and for scalar B3 
in the late phase. For B1 and B2, the best-fit timescale $\tau_{\rm con}$ is, 
respectively, 0.40 and 0.41 $\tau_{\rm dyn}$. For the early phase of scalar B3, 
we used the same nonlinear integral model as in Fig.\ \ref{fig:pfB1} with 
$\tau_{\rm int} = 0.3 \tau_{\rm dyn}$.  The late phase is fit by eq.\ (\ref{pfnconvsolution}) 
with $\tau_{\rm con}  = 0.42 \tau_{\rm dyn}$ and $n=4.6$. We combined the two 
phases at $P(Z_{\rm c}, t) =0.5$.  A comparison of Fig.\ \ref{fig:pfB1} and Fig.\ \ref{fig:pfB2} shows that, 
with eq.\ (\ref{pfnconvsolution}) from the generalized convolution model, 
the fitting quality is significantly improved.  
 
We point out that eq.\ (\ref{pfnconvsolution}) is used simply as a fitting function. 
The generalized convolution model (\S 3.4) behind this equation does not 
address the effects of shocks and compressibility on the PDF of passive scalars in supersonic 
turbulence. There is thus no physical reason why it provides successful fits to 
the pristine mass fraction in the $M=6.2$ flow. A physical closure model is motivated to 
successfully explain and match the evolution of $P(Z_{\rm c}, t)$ in highly supersonic 
turbulence. 

  
\section{Conclusions}

Motivated by the process of primordial star formation 
in the first generation of galaxies, we investigated the 
general problem of how the unpolluted flow material in 
compressible turbulence is contaminated by mixing. 
We approached this problem using both theoretical 
modeling and numerical simulations. The theoretical 
approach is based on the probability distribution 
method for turbulent mixing,  since the fraction of 
the unpolluted or slightly polluted mass corresponds 
to the left tail of the concentration PDF. We first derived 
an equation for the concentration PDF with density-weighting, 
where the advection term exactly conserves the global PDF. 
We then considered several existing closure models for the diffusivity term 
in the PDF equation, including the mapping closure 
model (Chen et al.\ 1989), the nonlinear integral 
models (Curl 1963, Dopazo 1979, Janicka et al.\ 1979) 
and the self-convolution models (Venaille and Sommeria 2007, 
Duplat and Villermaux 2008), and derived the predictions 
of these models for the exactly unpolluted fraction, $P(t)$, 
or for the fraction, $P(Z_{\rm c}, t)$, of the flow with 
$Z$ below a small threshold, $Z_{\rm c}$.  

To test and constrain the model predictions, we carried 
out numerical simulations evolving decaying scalars 
in two isothermal turbulent flows with rms Mach 
numbers of 0.9 and 6.2. Three passive scalars were 
included in each flow, and their initial pollutant fractions, 
$P_1$ were set to be 0.5, 0.1 and 0.01, respectively. 
We found that the mapping closure model gives satisfactory 
predictions for the central part and the high-$Z$ tails 
of the scalar PDF, but underestimates the low-$Z$ tail at 
large times, especially for scalars with small initial pollutant 
fraction. 
The left PDF tails become broader with increasing flow Mach number, 
and thus the discrepancy between the mapping closure prediction and 
the simulation results is larger at Mach 6.2. 
We showed that, in the $M=0.9$ flow, the scalar PDF is well fit by 
Gamma distributions at late times, as predicted by Villermaux and 
Duplat (2003). All the closure models adopted in our study were 
originally developed for mixing in incompressible turbulence, and they do not 
provide successful predictions for the scalar PDF in the highly 
supersonic flow.   

Our simulation results for $P(Z_{\rm c}, t)$ in the Mach 0.9 flow 
can be well fit by using eqs.\ (\ref{pfintegralsolution}) 
and (\ref{pfcconvsolution}) from the nonlinear integral model 
and the continuous convolution model of Venaille and Sommeria 
(2007), respectively. Although these two equations were 
originally derived for the fraction of exactly pollutant-free mass, 
they provide useful fitting functions for $P(Z_{\rm c}, t)$ with
a small finite threshold, $Z_{\rm c}$. We showed that, for 
the two scalars with $P_1 \ge 0.1$, the evolution of $P(Z_{\rm c}, t)$ 
follows the equation $\dot P(Z_{\rm c}, t) = P(Z_{\rm c}, t)\ln[P(Z_{\rm c}, t)]/\tau_{\rm con}$ 
from the continuous convolution model. On the the hand,  
for the scalar with $P_1 =0.01$, $P(Z_{\rm c}, t)$ 
shows different behaviors at early and late times. 
In the early phase, the evolution of $P(Z_{\rm c}, t)$ 
is consistent with the equation $\dot P(Z_{\rm c}, t) = -P(Z_{\rm c}, t)[1-P(Z_{\rm c}, t)]/\tau_{\rm int}$ 
from the nonlinear integral model, and the later phase is well fit by the continuous 
convolution model.  A satisfactory fit to the entire behavior of $P(Z_{\rm c}, t)$ 
was obtained by connecting the two phases. The continuous convolution 
model starts to apply once the polluted mass fraction is larger than 
0.1-0.2.

When normalized to the flow dynamical time ($\tau_{\rm dyn}$), 
the decay of $P(Z_{\rm c}, t)$ is slower in the $M=6.2$ for two 
reasons. First, the mixing timescale in units of $\tau_{\rm dyn}$,  
is about 20\% larger than in the $M=0.9$ flow. Second, 
at the same variance, the left tail of the scalar PDF is broader at 
higher Mach numbers. Due to the second effect, the shape of the  $P(Z_{\rm c}, t)$ 
curve as a function of time changes as the Mach number 
increases. We find that a generalized version of the self-convolution 
model ( \S 3.4 and \S 4.3; Duplat \& Villermaux 2008) 
provides a good fitting function, eq.\ (\ref{pfnconvsolution}), to the 
evolution of $P(Z_{\rm c}, t)$ in highly supersonic turbulence. 
With $n=4.6$, this equation matches our simulation data well for the two 
scalars with $P_1 \ge 0.1$. Like the case of the $M=0.9$ flow, 
we obtained a good fit to the simulation result for the scalar with 
$P_1 =0.01$ by combining different behaviors at early and late times.  At early times, 
we used eq.\ (\ref{pfintegralsolution}) from the nonlinear integral 
model, while the later phase was fit by eq.\ (\ref{pfnconvsolution}) with 
$n=4.6$. We point out that, although it provides a good fitting 
function to the pristine mass fraction, the generalized convolution 
model does not have a physical connection to how the flow 
compressibility affects turbulent mixing in supersonic flows. Physical PDF closure models 
are motivated to directly incorporate the effects of shocks or the flow 
compressibility on mixing in supersonic turbulence.   
    
The fitting functions obtained this study can be used 
to develop a subgrid model for large-scale simulations for 
mixing of heavy elements in the interstellar media of 
early galaxies. Such a subgrid model would provide an important step 
toward predicting the fraction of pristine gas in the first generation 
of galaxies. In order to apply our results with higher 
accuracy, we will carry out a systematic numerical study in a future work covering a 
broader range of parameters including varying initial scalar 
conditions and turbulent Mach numbers.  


\acknowledgements

L.P. thanks Prof. Guo-Wei He for helpful discussions on the mapping closure model, and the Institute of Mechanics, Chinese 
Academy of Sciences for its hospitality during his visit.  L.P. and E.S. acknowledge support from NASA under theory 
Grant No. NNX09AD106 and astrobiology institute grant 08-NAI5-0018 and  from the National Science Foundation 
under grant AST 11-03608. J.S. acknowledges support by the NASA Astrobiology Institute, Virtual Planetary Laboratory
Lead Team. All simulations were conducted at the Arizona State University Advanced Computing Center and the Texas 
Advanced Computing Center, using the FLASH code, a product of the DOE ASC/Alliances-funded Center for Astrophysical 
Thermonuclear Flashes at the University of Chicago.

\appendix

\section{The PDF equation}

In this appendix, we derive the equation for the concentration PDF in compressible 
flows using the technique developed by Lundgren (1967).  
Similar derivations for the scalar PDF equation can be 
found in, e.g., Pope (1976), O'Brien (1980), 
Dapazo et al.\ (1997) and Pope (2000). 
Unlike these previous derivations, we adopt a 
density-weighting scheme, which is needed for 
passive scalar mixing in compressible turbulence.

Our derivation makes use of a statistical ensemble consisting of many 
independent realizations. We start with the definition of the fine-grained 
PDF in a single realization. Because in a specific realization 
the concentration field is single-valued at given 
position (${\bf x}$) and time ($t$), 
the fine-grained PDF is a delta function
\begin{equation}
q'(Z;{\bf x},t)= \delta [Z-C({\bf x},t)],
\label{volumefinepdf}
\end{equation}
where $Z$ is the sampling variable. Considering the existence 
of significant density fluctuations in supersonic flows, 
we define a fine-grained PDF with density-weighting
\begin{equation}
p'(Z;{\bf x},t)= \tilde{\rho} ({\bf x},t) \delta [Z-C({\bf x},t)], 
\label{massfinepdf}
\end{equation}
where the density-weighting factor $\tilde{\rho}$ is the ratio of the 
local flow density ${\rho}({\bf x},t)$ to the average density $\bar{\rho}$. 
These two fine-grained PDFs are functions of $Z$, ${\bf x}$ and $t$, 
and their dependence on space and time is though $C({\bf x},t)$ 
and $\rho({\bf x},t)$.  

We calculate the time-derivatives of $q'(Z;{\bf x},t)$ and 
$p'(Z;{\bf x},t)$. Since $q'(Z;{\bf x},t)$ depends on $t$ 
only through the quantity $Z-C({\bf x},t)$, we have,  
\begin{equation}
\frac {\partial q'(Z;{\bf x},t)}{\partial t}  = - \frac {\partial C({\bf x},t) } {\partial t} \frac {\partial q'(Z;{\bf x},t)} 
{\partial Z}. 
\label{volumepdfderivative} 
\end{equation}
Using  eqs.\ (\ref{massfinepdf}) and (\ref{volumepdfderivative}), 
we obtain the time-derivative of $p'(Z;{\bf x},t)$,  
\begin{equation}
\frac {\partial p'(Z;{\bf x},t)}{\partial t} = \frac {\partial \tilde{\rho}}{\partial t} q'(Z;{\bf x},t)- 
\tilde{\rho} \frac {\partial C({\bf x},t)}{\partial t} \frac {\partial q'(Z;{\bf x},t)}{\partial Z}. 
\label{timederivativefinal}
\end{equation}
Similarly, we can derive an equation for ${\bf \nabla}\cdot(p' {\bf v} )$,    
\begin{equation}
{\bf \nabla} \cdot(p' {\bf v} ) =  \left[ \nabla \cdot (\tilde{\rho} {\bf v}) \right] q'(Z;{\bf x},t)
- \tilde{\rho} {\bf v} \cdot {\bf \nabla} C({\bf x},t) \frac {\partial q'(Z;{\bf x},t)}{\partial Z}.   
\label{spatialderivative}
\end{equation}

We add eqs.\ (\ref{timederivativefinal}) and (\ref{spatialderivative}). 
From the continuity equation for $\tilde{\rho}({\bf x},t)$, the first terms on 
the right hand sides of eqs.\ (\ref{timederivativefinal}) and (\ref{spatialderivative}) 
add up to zero. Using the advection-diffusion equation (\ref{advection}) of 
$C({\bf x},t)$ for the sum of the last terms in these two equations, we find,  
\begin{equation}
\frac {\partial p'(Z;{\bf x},t)}{\partial t} + 
{\bf \nabla} \cdot (p'(Z;{\bf x},t) {\bf v} ) = - \frac{\partial}{\partial Z} \left[p'(Z;{\bf x},t) \left(\frac{1}{\rho} \nabla \cdot (\rho \kappa \nabla C)+S\right) \right],
\label{finegraineq}
\end{equation}
where we used the fact that, except $p'(Z;{\bf x},t)$ or $q'(Z;{\bf x},t)$, 
all the quantities in the right-hand-side term are independent of $Z$. 
Eq.\ (\ref{finegraineq}) is essentially a Liouville equation. 
In analogy to the kinetic theory, the concentration here corresponds to the particle 
momentum, and the equation 
$dC/dt= \frac{1}{\rho}\nabla \cdot (\rho \kappa \nabla C)+S$ 
corresponds to the particle equation of motion. 

We now consider the coarse-grained PDF, defined as 
the ensemble average of the fine-grained PDFs 
over independent realizations. The coarse-grained PDFs with 
volume- and density-weighting are, respectively,  
$q(Z;{\bf x},t)=\langle q'(Z;{\bf x},t)\rangle$ and 
$p(Z;{\bf x},t)=\langle p'(Z;{\bf x},t)\rangle$, where 
$\langle\cdot\cdot\cdot\rangle$ denotes the ensemble average. 
The average is over different values of the concentration, 
$C({\bf x},t)$, the flow velocity and density, $\bf {v} ({\bf x},t)$ 
and $\rho ({\bf x},t)$, and the scalar source, $S({\bf x},t),$ in different realizations.     
From eq.\ (\ref{finegraineq}), it immediately follows that, 
\begin{equation}
\frac {\partial p(Z;{\bf x},t)}{\partial t}   
+ {\bf \nabla} \cdot \langle p'(Z;{\bf x},t) {\bf v} \rangle 
= -\frac {\partial}{\partial Z} \left\langle p'(Z;{\bf x},t) \left(\frac{1}{\rho}\nabla \cdot (\rho \kappa \nabla C)+S\right)\right\rangle.
\label{coarsepdfeq}
\end{equation} 

The ensemble average terms in eq.\ (\ref{coarsepdfeq}) 
can be expressed in more convenient forms. 
For  any stochastic quantity $B({\bf x},t)$, 
we have the following relation (see, e.g., Pope 2000)
\begin{equation}
\langle q'(Z;{\bf x},t) B({\bf x},t) \rangle = q(Z;{\bf x},t) \langle B({\bf x},t)|C({\bf x},t)=Z \rangle,
\label{conditionalrelation}
\end{equation}
where $\langle B({\bf x},t)|C({\bf x},t) = Z\rangle$ denotes the ensemble average 
of $B({\bf x},t)$ conditioned on $C({\bf x},t)=Z$. The conditional mean appears 
here because the factor $q'(Z;{\bf x},t)$ selects only the realizations 
where $C({\bf x}, t)$ is equal to $Z$. Using eq.\ (\ref{conditionalrelation}), we have 
\begin{equation}
\langle p'(Z;{\bf x},t) B({\bf x},t) \rangle = p(Z;{\bf x},t) \frac {\langle \rho B|C({\bf x},t)=Z \rangle}
{\langle \rho |C({\bf x},t)=Z \rangle},
\label{conditional}
\end{equation}
where we used $p(Z;{\bf x},t)=q(Z;{\bf x},t)\langle \tilde{\rho}|C({\bf x},t)=Z\rangle$. 

With eqs.\ (\ref{coarsepdfeq}) and (\ref{conditional}), we arrive at the final equation 
for the coarse-grained PDF with density weighting,  
\begin{equation}
\frac {\partial p (Z; {\bf x},t) }{\partial t} +  {\bf \nabla} \cdot \left(p \frac {\langle \rho {\bf v}|C=Z \rangle}
{\langle \rho |C=Z \rangle} \right)
= - \frac {\partial}{\partial Z} \left( p \frac {\langle \nabla \cdot 
(\rho \kappa \nabla C)|C=Z \rangle}{\langle \rho|C=Z \rangle} \right) -
\frac {\partial}{\partial Z} \left( p \frac {\langle \rho S|C=Z\rangle}  {\langle \rho|C=Z \rangle} \right),
\label{longform}
\end{equation}
where we replaced the condition $C({\bf x},t)=Z$ by $C=Z$ 
for the simplicity of notations. Note that the advection term 
is in a divergence form, which is the motivation for the 
use of a density-weighting scheme in our derivation. 
A detailed physical discussion of all the terms in this equation is given 
in the text. 

 
\bibliographystyle{jfm}
\bibliography{ms}

\end{document}